\documentclass[12pt]{article}
\usepackage{cite}
\usepackage{graphicx}
\usepackage{multicol}
\usepackage{amsfonts}
\usepackage{amssymb}
\usepackage{amsmath}
\usepackage{heck}
\usepackage{setspace}
\usepackage{verbatim}
\usepackage{color}
\usepackage{epsfig}
\usepackage[margin=1in]{geometry}

\providecommand{\U}[1]{\protect\rule{.1in}{.1in}}
\numberwithin{equation}{section}

\newcommand{\ba}{\begin{eqnarray}}
\newcommand{\ea}{\end{eqnarray}}

\begin{document}

\date{December 2012}

\title{$S$ and $T$ for SCFTs}

\institution{HARVARD}{\centerline{${}^{1}$Jefferson Physical Laboratory, Harvard University, Cambridge, MA 02138, USA}}

\institution{YALE}{\centerline{${}^{2}$Department of Physics, Yale University, New Haven, CT 06520, USA}}

\institution{QUEEN}{\centerline{${}^{3}$Centre for Research in String Theory, Queen Mary, University of London, London E1 4NS, UK}}

\authors{Jonathan J. Heckman\worksat{\HARVARD}\footnote{e-mail: {\tt jheckman@physics.harvard.edu}},
Piyush Kumar \worksat{\YALE}\footnote{e-mail: {\tt piyush.kumar@yale.edu}}, and
Brian Wecht \worksat{\QUEEN}\footnote{e-mail: {\tt b.wecht@qmul.ac.uk}}}

\abstract{
We develop a general set of methods for computing the
oblique electroweak parameters $S$ and $T$ for bottom-up and top-down motivated scenarios
in which the Higgs weakly mixes with a superconformal extra sector. In addition to
their utility in phenomenological studies, the observables $S$ and $T$ are also of purely theoretical
interest as they are defined by correlation functions of broken symmetries.
We show that in the limit where the extra sector
enjoys an approximate custodial symmetry, the leading
contributions to $S$ and $T$ can be recast as calculable data of the theory in the conformal
phase. Using this result, we also obtain model-independent
bounds on the sign and size of oblique electroweak corrections
from unitary superconformal theories.}

\maketitle

\enlargethispage{\baselineskip}

\setcounter{tocdepth}{2}

\tableofcontents

\section{Introduction}

The discovery of a Higgs-like resonance near 125 GeV at the LHC \cite{Aad:2012gk, Chatrchyan:2012gu} signals
the start of a dynamic new phase in particle theory and experiment. Needless to say, measuring
the detailed properties of this resonance will provide crucial insights into the
nature of electroweak symmetry breaking as well as possible clues to beyond-the-Standard
Model (BSM) physics.

By far the simplest and best motivated interpretation is that this resonance is
indeed a Higgs boson. There are two ways in which properties of the Higgs can
provide information about BSM\ physics. First, the measured mass of the Higgs
implies that the quartic coupling in the Higgs potential is not very small.
This provides a simple way to discriminate between various models of
electroweak symmetry breaking. Additionally, BSM physics, especially in
\textquotedblleft natural\textquotedblright\ theories, will modify the Higgs
self-couplings as well as couplings to SM gauge bosons and fermions in general, thereby
affecting production and decay modes.

These two effects are intricately related in natural theories. For example, in
the minimal supersymmetric Standard Model (MSSM), the quartic coupling at tree
level is determined by the $SU(2)$ and $U(1)_{Y}$ gauge couplings, and is
extremely small. Hence, additional contributions to the quartic coupling are
needed. A natural way to allow for such contributions (\textit{i.e.}, without
introducing fine tuning in the Higgs potential) is by coupling the Higgs to
states beyond the MSSM. Since these states couple to the Higgs, they can also
change production and decay of the Higgs relative to the SM
and MSSM, for example via the dimension-five operators for gluon-fusion
production ($gg\rightarrow h^{0}$) and diphoton decay ($h^{0}\rightarrow
\gamma\gamma$) \cite{Shifman:1979eb}.

Although in principle one could imagine a wide variety of states beyond the SM
(or MSSM), precision electroweak measurements impose constraints on such extra
sectors. In particular, the parameters known as $S$ and $T$ are the dominant
contributions in the oblique approximation \cite{Peskin:1990zt, Peskin:1991sw}%
. These parameters describe certain dimension-six operators that can appear in
the effective theory below the scale of new physics. Present constraints on
$S$ and $T$ allow for order $0.1$ deviations at the $68\%$ confidence
level \cite{ErlerLangackerPDG}. Contributions beyond this level are already
difficult to accommodate. Indeed, even before the direct observation of a
Standard Model-like Higgs, precision electroweak fits already indicated a good
fit for the hypothesis of a weakly coupled Standard Model Higgs near $100$ GeV
\cite{ALEPH:2005ab}.

Independent of its connection to phenomenology, the calculation of the
parameters $S$ and $T$ is also quite interesting for purely formal reasons.
This is because these observables involve correlation functions of
\textit{broken} symmetry generators:%
\begin{align}
S &  =-16\pi\,\frac{\Pi_{3Y}(M_{Z}^{2})-\Pi_{3Y}(0)}{M_{Z}^{2}}\label{Sdef}\\
\,T &  =4\pi\,\frac{\Pi_{11}(0)-\Pi_{33}(0)}{s_{W}^{2}c_{W}^{2}M_{Z}^{2}%
}\label{Tdef}%
\end{align}
where $s_{W}$ and $c_{W}$ stand for $\sin\theta_{W}$ and $\cos\theta_{W}$,
respectively, and $\Pi_{ij}(q^{2})$ denotes the scalar factor of the vacuum
polarization amplitude:%
\begin{equation}
\langle J_{i}^{\mu}(q)J_{j}^{\nu}(-q)\rangle=i\eta^{\mu\nu}\,\Pi_{ij}%
(q^{2})+q^{\mu}q^{\nu}\,\mathrm{terms}\label{Pi-def}%
\end{equation}
for Fourier transformed symmetry currents $J_{i}^{\mu}$ evaluated at a
reference four-momentum $q^{\mu}$. The subscripts on $\Pi_{ij}$ and
$J_{i}^{\mu}$ refer to either the components of weak $SU(2)$ (for $i=1,2,3$)
or the hypercharge (for $i=Y$). Polarization amplitudes for \textit{unbroken}
symmetry generators are highly constrained by gauge invariance. Moreover, in
supersymmetric gauge theories, it is often possible to calculate the leading
behavior of these amplitudes by relating them to the computation of the NSVZ
beta function \cite{Novikov:1983uc}. The contributions $\Delta S$ and $\Delta
T$ from an extra sector are even more informative as they are sensitive to
details of symmetry breaking and the mass spectrum of new states.

In spite of their important role in both formal theory and phenomenology, it
has proven notoriously difficult to extract quantitative estimates for $\Delta
S$ and $\Delta T$ in strongly coupled extra sectors. This is because such
computations appear to require detailed knowledge of the mass spectrum. For
example, in order to estimate $\Delta S$ in technicolor theories, one has to
make assumptions about the spectrum of strongly coupled bound states. One
possibility is to assume the UV behavior resembles that of QCD, although other
possibilities such as \textquotedblleft walking\textquotedblright\ in an
approximate conformal phase can also occur.

In this paper, we develop a set of techniques to compute $\Delta S$ and
$\Delta T$ when the Higgs sector mixes with an additional sector which becomes
superconformal at high energies. From the perspective of the extra sector, we
will take the gauge group of the Standard Model to be a flavor symmetry. We
couple the supersymmetric two Higgs doublet sector $H_{u}\oplus H_{d}$ to
operators $\mathcal{O}_{u}$ and $\mathcal{O}_{d}$ of the extra sector via:%
\begin{equation}
\mathcal{L}_{eff}\supset\int d^{2}\theta\text{ }\left(  \lambda_{u}%
H_{u}\mathcal{O}_{u}+\lambda_{d}H_{d}\mathcal{O}_{d}\right)  +h.c..
\label{perts}%
\end{equation}
Scenarios involving such mixing terms have been considered in
\cite{Stancato:2008mp, Komargodski:2008ax, Azatov:2011ht, Azatov:2011ps, Gherghetta:2011na, DSSM,
HoloHiggs, Kitano:2012wv}.

The bottom-up motivation for this class of couplings is that compared with the
MSSM, they can raise the mass of the lightest Higgs \cite{DSSM}, and also
allow a broader class of mixing angles in the (supersymmetric)
two-Higgs-doublet model (s2HDM). The simplest way to generate a large
correction to the Higgs mass is by increasing the size of the Yukawas
$\lambda_{u}$ and $\lambda_{d}$, which also pushes the mass of the extra
states higher than the weak scale. To avoid issues with low-scale Landau
poles, it is therefore natural to envisage a situation in which the extra
sector enters an approximate (super)conformal phase at higher energies.

Such extra sectors are also well-motivated in various string constructions. For example,
they can arise from D3-branes probing a stack of intersecting seven-branes \cite{Funparticles}.
These D3-brane sectors also typically couple to the third generation of MSSM states, providing a simple
mechanism to alter the masses of third generation superpartners relative to
first and second generation superpartners \cite{HVW}. From this perspective,
constraints on $S$ and $T$ provide valuable information on possible stringy
extensions of the Standard Model.

The phenomenology of such extra sectors is potentially quite rich, though also
very model dependent. In principle, extra sector states could be charged under
just $SU(2)_{W}\times U(1)_{Y}$, or may also include colored states as well.
There could also be hierarchies between the colored and uncolored states, as
well as additional gauge singlets which could facilitate hard-to-detect decay
modes. Our aim here will be to derive constraints which do not involve such
highly model dependent details. Combining this with earlier work on the
dimension-five operators responsible for gluon fusion production and diphoton decays
\cite{HoloHiggs}, this provides a more complete picture of the low energy
consequences of mixing between the Higgs and a superconformal extra sector.

The basic outline of the present work will be to study the contributions to
$S$ and $T$ when the Higgs vevs are the sole source of mass for the states of
the extra sector. Of course, in realistic model building applications it is
often necessary to introduce an additional vector-like mass for states of the
extra sector, in which case there is an overall suppression by a factor of
$v^{2}/\Lambda_{vec}^{2}$. To evaluate the leading contributions to $S$ and
$T$ in the absence of such vector-like masses we shall approximate the
correlation functions by two-point functions evaluated in the superconformal
phase. To this end, we shall assume that there is an approximate custodial
symmetry so that the flavor symmetry of the extra sector is $SU(2)_{W}\times
SU(2)_{\widetilde{W}}\times U(1)_{\chi}$, with a custodial $SU(2)_{diag}%
\subset SU(2)_{W}\times SU(2)_{\widetilde{W}}$, where $SU(2)_{W}$ denotes the
weak sector gauge group (this symmetry could, however, be broken by the Higgs
vevs). In this language, the hypercharge $Y$ is given by $Y=T_{\widetilde{W}%
}+T_{\chi}$, where $T_{\widetilde{W}}$ denotes the $\sigma_{(3)}$ generator in
$SU(2)_{\widetilde{W}}$. The assumption of a custodial $SU(2)$ appears to be
fairly mild, since we know $T$ to be small. Using this, we characterize the
contributions to both the $S$ and $T$ parameters in terms of calculable
quantities in the conformal theory.

We show that the calculation of $\Delta T$ can be recast as a computation of
wave-function renormalization of the Higgs field K\"{a}hler potential
\cite{DSSM}. The resulting contribution to $\Delta T$ can be expressed
as a change in the ratio of the W- and Z-boson masses, leading to the expression:
\begin{equation}
\Delta T=\frac{1}{\alpha_{EM}}\cos^{2}2\beta\times\delta
\end{equation}
where $\tan\beta=v_{u}/v_{d}$ denotes the ratio of the Higgs vevs and
$\delta=\Delta-1$ denotes the excess dimension of the Higgs, viewed as a field
of the conformal sector. The scaling dimension of the Higgs can often be
extracted by determining the infrared R-symmetry of the superconformal theory,
via a-maximization \cite{Intriligator:2003jj}.

The calculation of $\Delta S$ involves additional details of the extra sector.
We find, however, that to leading order in the operator perturbations of
(\ref{perts}), this contribution is independent of the details of the mass
spectrum and is given by:%
\begin{equation}
\Delta S_{(0)}=\frac{b_{AA}}{4\pi}%
\end{equation}
where $b_{AA}$ is a \textit{calculable} anomaly coefficient:%
\begin{equation}
b_{AA}=-3\text{Tr}(R_{IR}J_{A}J_{A}).
\end{equation}
Here, $R_{IR}$ is the R-symmetry of the CFT, and $J_{A}$ is the axial current
$J_{W}-J_{\widetilde{W}}$, formed from the $U(1)_{W}\times U(1)_{\widetilde
{W}}$ subalgebra of $SU(2)_{W}\times SU(2)_{\widetilde{W}}$.

With sufficient additional global symmetries, we can say even more, and
compute subleading contributions to $\Delta S$ which are sensitive to the mass
spectrum of the theory. In situations where the states of the extra sector get
their mass dominantly from either $H_{u}$ or $H_{d}$, but not both, the flavor
symmetry is effectively doubled, leaving us with a different $U(1)_{W}\times
U(1)_{\widetilde{W}}\times U(1)_{\chi}$ acting on each set of states
independently. From a bottom-up perspective, this assumption is
well-motivated, as it suppresses contributions to
flavor changing neutral currents in a two Higgs doublet model. From a top-down
perspective, this assumption is also satisfied in SQCD-like models and some
string/F-theory constructions; we will describe this further in section
\ref{sec:EXAMPLES}. We denote the \textquotedblleft up-type" symmetries by
$W_{u},\widetilde{W}_{u},\chi_{u}$, and similarly for the \textquotedblleft
down-type" symmetries. In this case, we can evaluate a further anomaly:%
\begin{equation}
b_{W_{u}Y_{u}}=-3\text{Tr}(R_{IR}J_{W_{u}}J_{Y_{u}}).
\end{equation}
Including this contribution allows us to take into account further
(logarthmic) contributions which are sensitive to the breaking of the
$SU(2)_{cust}$ symmetry:%
\begin{equation}
\Delta S=\frac{b_{AA}}{4\pi}-\frac{4b_{W_{u}Y_{u}}}{2\pi}\log\frac{m_{u}%
}{m_{d}}%
\end{equation}
where $m_{u}=\lambda_{u}v_{u}$ and $m_{d}=\lambda_{d}v_{d}$ define
characteristic mass scales for the up- and down-type states.

Even without calculating the precise values of $\Delta S$ and $\Delta T$ from
a conformal sector, we can still extract valuable model-independent
information about the overall sign and relative size of the various
contributions to these parameters. We find that in unitary theories where the
Standard Model gauge groups remain weakly coupled, the sign of $\Delta T$ is
fixed, and is non-negative. Moreover, we find that the non-logarithmic
contribution $\Delta S_{(0)}\geq0$. A priori, however, $\Delta S$ can have either sign
depending on the overall hierarchy between $m_{u}$ and $m_{d}$.

The rest of this paper is organized as follows. First, in section
\ref{sec:REVIEW} we review some general properties of $S$ and $T$, and their
evaluation in weakly coupled models. The striking simplicity of these results
motivates our expectation that a similar result is available for more general
extra sectors. In section \ref{sec:SETUP} we discuss our setup in more detail,
and show in sections \ref{sec:TPARAM} and \ref{sec:SPARAM} how to calculate
the leading oblique electroweak corrections from a superconformal extra sector.
In section \ref{sec:UNIT} we determine some model-independent bounds on the
sign and size of $\Delta S$ and $\Delta T$, and in section
\ref{sec:EXAMPLES}\ we present some example computations. Section
\ref{sec:CONC} contains our conclusions.

\section{Review of Precision Electroweak Parameters \label{sec:REVIEW}}

In this section we review the definitions of the oblique electroweak
parameters $S$ and $T$. In preparation for our later discussion, we also
present a schematic calculation of $S$ and $T$ for weakly coupled states. In
subsequent sections we will aim to show how these contributions generalize to
a (possibly strongly coupled) extra sector.

Let us begin with the definitions of the oblique electroweak parameters. In
extensions of the Standard Model, new states can in principle alter the masses
$M_{W}$ and $M_{Z}$ of the W- and Z-bosons, as well as the value of the Fermi
decay constant $G_{F}$, the weak mixing angle $\theta_{W}$, and the gauge
couplings $g$ and $g^{\prime}$ for $SU(2)_{W}$ and $U(1)_{Y}$. An efficient
way to parameterize such deviations is in terms of precision electroweak
observables, connected with vacuum polarization amplitudes for the electroweak
sector. We will assume throughout this work that the oblique approximation is
valid, meaning that the primary contributions to precision observables are
captured by $S$ and $T$. These are defined by equations (\ref{Sdef}) and
(\ref{Tdef}), which we reproduce here for convenience of the reader:%
\begin{align}
S  &  =-16\pi\,\frac{\Pi_{3Y}(M_{Z}^{2})-\Pi_{3Y}(0)}{M_{Z}^{2}}\\
\,T  &  =4\pi\,\frac{\Pi_{11}(0)-\Pi_{33}(0)}{s_{W}^{2}c_{W}^{2}M_{Z}^{2}}.
\end{align}
In the limit where $M_{Z}$ is much smaller than the characteristic mass of new
states, $S$ is specified by the derivative of $\Pi_{3Y}$ with respect to
$q^{2}$, the square of the momentum transfer:%
\begin{equation}
S=-16\pi\,\Pi_{3Y}^{\prime}(0). \label{S-def}%
\end{equation}
In this sense, $S$ and $T$ involve the IR\ behavior of the vacuum polarization amplitudes.

Our primary interest is in contributions from new physics, which we denote by
$\Delta S$ and $\Delta T$. In the models of interest to us here, the extension
consists of two portions. First, there are the possible contributions from
extending the Standard Model to the MSSM. This includes contributions from
extending the Higgs sector to a supersymmetric two Higgs doublet model
(s2HDM), as well as the other superpartners of the MSSM. Additionally, there
are the contributions from the extra sector. In the limit where the Higgs
sector only weakly mixes with an extra sector, we can treat its contribution
to $S$ and $T$ independently from the s2HDM and MSSM contributions.

After the states of the extra sector pick up a mass of order $\Lambda$, we can
integrate them out. For example, the higher dimension operators which
contribute to $S$ and $T$ are:%
\begin{align}
\mathcal{O}_{S}  &  =\left(  c_{S}^{(u)}\frac{H_{u}^{\dag}\sigma_{(i)}H_{u}%
}{\Lambda^{2}}+c_{S}^{(d)}\frac{H_{d}\sigma_{(i)}H_{d}^{\dag}}{\Lambda^{2}%
}+c_{S}^{(mix)}\frac{H_{d}\sigma_{(i)}H_{u}+H_{u}^{\dag}\sigma_{(i)}%
H_{d}^{\dag}}{\Lambda^{2}}\right)  W_{\mu\nu}^{(i)}B^{\mu\nu}\label{OS}\\
\mathcal{O}_{T}  &  =c_{T}^{(u)}\frac{\left\vert H_{u}^{\dag}D_{\mu}%
H_{u}\right\vert ^{2}}{\Lambda^{2}}+c_{T}^{(d)}\frac{\left\vert H_{d}^{\dag
}D_{\mu}H_{d}\right\vert ^{2}}{\Lambda^{2}}+c_{T}^{(mix)}\frac{\left\vert
H_{u}D_{\mu}H_{d}\right\vert ^{2}}{\Lambda^{2}}%
\end{align}
where $\frac{\sigma_{(i)}}{2}W_{\mu\nu}^{(i)}$ denotes the $SU(2)$ field
strength and $B_{\mu\nu}$ denotes the $U(1)_{Y}$ field strength. Once the
Higgs fields develop a non-zero vev, we obtain non-zero contributions to the
gauge boson two point functions entering the definitions of $S$ and $T$. In
other words, our task is to compute the \textquotedblleft order one
coefficients\textquotedblright\ denoted by the $c$'s.

In a weakly coupled setting, the contributions to $S$ and $T$ from extra
states are fully calculable. For examples of such calculations, see
\cite{Peskin:1990zt, Peskin:1991sw, Dugan:1991ck, Gates:1991uu, He:2001tp},
and \cite{Skiba:2010xn} for a pedagogical discussion. Consider a weakly
coupled left-handed fermion doublet $\left(  \psi_{1},\psi_{2}\right)  $ with
hypercharge $Y$, and its $SU(2)_{W}$ singlet partners $\psi_{1}^{c}$ and
$\psi_{2}^{c}$, with masses $m_{1},m_{2}$. These states pick up a mass once
the Higgs fields develop vevs. In our conventions, $H_{u}$ and $H_{d}$ have
$U(1)_{Y}$ hypercharge $+1/2$ and $-1/2$, respectively. The contributions
$\Delta S$ and $\Delta T$ are then (see e.g. \cite{He:2001tp}):%
\begin{align}
\Delta S &  =\frac{1}{6\pi}\left(  1-2Y\log\frac{m_{1}^{2}}{m_{2}^{2}}\right) \label{Sweak}
\\
\Delta\,T &  =\frac{1}{8\pi s_{W}^{2}c_{W}^{2}M_{Z}^{2}}\left(  \frac
{m_{1}^{2}+m_{2}^{2}}{2}-\frac{m_{1}^{2}m_{2}^{2}}{m_{1}^{2}-m_{2}^{2}}%
\log\frac{m_{1}^{2}}{m_{2}^{2}}\right)  .\label{Tweak}%
\end{align}

Particularly suggestive is the contribution $\Delta S$ for a collection of
Dirac fermions $\psi$ of mass $m_{(\psi)}$ with weak isospin $t_{L}^{(\psi
)},t_{R}^{(\psi)}$ and hypercharge $y_{L}^{(\psi)},y_{R}^{(\psi)}$ for the
left- and right-handed components:\footnote{We thank P. Langacker for helpful
discussions.}%
\begin{equation}
\Delta S=\frac{1}{3\pi}\underset{\psi}{%
%TCIMACRO{\dsum }%
%BeginExpansion
{\displaystyle\sum}
%EndExpansion
}\left[  \left(  t_{L}^{(\psi)}-t_{R}^{(\psi)}\right)  ^{2}-2\left(
t_{L}^{(\psi)}y_{L}^{(\psi)}+t_{R}^{(\psi)}y_{R}^{(\psi)}\right)  \log
\frac{m_{(\psi)}^{2}}{\mu_{(0)}^{2}}\right]  \label{weakS}%
\end{equation}
with $\mu_{(0)}$ a UV cutoff. Since the total $t_{L}^{(\psi)}y_{L}^{(\psi)}$
and $t_{R}^{(\psi)}y_{R}^{(\psi)}$ vanish for each $SU(2)_{W}$ multiplet,
$\Delta S$ is independent of $\mu_{(0)}$, as it must be for a finite contribution.

One way to understand this result is to work in terms of a weakly coupled
basis of left-handed Weyl-fermions $\psi_{L}$ and $\psi_{R}^{c}$ with
effective Lagrangian:%
\begin{equation}
L_{(\psi)}=-i\psi_{L}^{\dag}\overline{\sigma}^{\mu}\partial_{\mu}\psi
_{L}-i\psi_{R}^{c\dag}\overline{\sigma}^{\mu}\partial_{\mu}\psi_{R}%
^{c}-m_{(\psi)}\psi_{L}\psi_{R}^{c}-m_{(\psi)}^{\dag}\psi_{L}^{\dag}\psi
_{R}^{c\dag}%
\end{equation}
where we have suppressed the implicit sum over the full $SU(2)_{W}$ multiplet,
since this symmetry is in general broken. Our main interest is the abelian
symmetries $U(1)_{W}\times U(1)_{Y}$, specified by weak isospin and
hypercharge. To account for the appearance of left-handed and (conjugate)
right-handed states, it is convenient to introduce an additional
$U(1)_{\widetilde{W}}$ symmetry which encodes the $U(1)_{W}$ charge of the
parity conjugate state. Each field of the theory then carries a specific
charge under $U(1)_{W}\times U(1)_{\widetilde{W}}\times U(1)_{\chi}$:
\begin{equation}%
\begin{tabular}
[c]{|c|c|c|c|}\hline
& $U(1)_{W}$ & $U(1)_{\widetilde{W}}$ & $U(1)_{\chi}$\\\hline
$\psi_{L}$ & $+t_{L}$ & $+t_{R}$ & $+\chi$\\\hline
$\psi_{R}^{c}$ & $-t_{R}$ & $-t_{L}$ & $-\chi$\\\hline
$m_{(\psi)}$ & $-(t_{L}-t_{R})$ & $-(t_{R}-t_{L})$ & $0$\\\hline
\end{tabular}
\end{equation}
where weak hypercharge is given by the generator $Y = T_{\widetilde{W}} + T_{\chi}$.
We can also introduce the vector and axial combinations
$T_{V}=T_{W}+T_{\widetilde{W}}$ and $T_{A}=T_{W}-T_{\widetilde
{W}}$. From this perspective, we see that the mass term is
charged under the axial $U(1)_{A}$, but neutral under
$U(1)_{V} \times U(1)_{\chi}$.

Evaluation of the $S$-parameter can now be understood by considering the
$\psi_{L}$ and/or $\psi_{R}^{c}$ running in the loop. There are two ways to
generate a non-zero contribution to $S$: either from a loop with both
$\psi_{L}$ and $\psi_{R}^{c}$ and a mass insertion, or a loop with either
$\psi_{L}$ or $\psi_{R}^{c}$ and no mass insertion.

It is straightforward to see that when only $\psi_{L}$ or $\psi_{R}^{c}$ run
in the loop, the contribution is proportional to $t_{L}^{(\psi)}y_{L}^{(\psi
)}$ or $t_{R}^{(\psi)}y_{R}^{(\psi)}$, times a logarithmic factor. This is the
second contribution in (\ref{weakS}).

To get a non-zero contribution when both $\psi_{L}$ and $\psi_{R}^{c}$ appear
in a loop, we must allow two mass insertions. Each mass insertion is neutral
under $U(1)_{V}$ but carries axial charge proportional to $t_{L}-t_{R}$. Thus,
the overall contribution from mass insertions is proportional to $\left(
t_{L}-t_{R}\right)  ^{2}$, as appropriate for the two-point function for the
axial current. This corresponds to the first term in equation (\ref{weakS}).

\section{Mixing with a Superconformal Extra Sector \label{sec:SETUP}}

In preparation for our discussion in subsequent sections, in this section we
provide more details on our setup, as well as the tools available for studying
SCFTs. As mentioned in the Introduction, our setup consists of a
supersymmetric two Higgs doublet sector which couples to a superconformal
extra sector via the F-terms:%
\begin{equation}
\mathcal{L}_{eff}\supset\int d^{2}\theta\text{ }\left(  \lambda_{u}%
H_{u}\mathcal{O}_{u}+\lambda_{d}H_{d}\mathcal{O}_{d}\right)
+h.c.\label{leffpert}%
\end{equation}
where $\mathcal{O}_{u}$ and $\mathcal{O}_{d}$ are operators of the extra
sector. See \cite{Stancato:2008mp, Azatov:2011ht, Azatov:2011ps,
Gherghetta:2011na, DSSM, HoloHiggs, Kitano:2012wv} for some examples of Higgs/extra sector mixing.

More precisely, we first imagine a \textquotedblleft UV theory" (close to the
GUT scale) specified by a superconformal fixed point, where the couplings
$\lambda_{u}$ and $\lambda_{d}$ have been switched off and the Higgs fields
have dimension one. To give an unambiguous meaning to the Yukawas $\lambda
_{u}$ and $\lambda_{d}$, we assume that the two-point functions for the
$\mathcal{O}$'s are normalized so that at the UV fixed point,
prior to mixing with the Higgs, we have:%
\begin{equation}
\left\langle \mathcal{O}^{\dag}(x)\mathcal{O}(0)\right\rangle =\frac{1}%
{16\pi^{4}}\frac{1}{x^{2\Delta_{\mathcal{O}}}}%
\end{equation}
where $\Delta_{\mathcal{O}}$ is the scaling dimension of $\mathcal{O}$.

We couple the Higgs to the extra sector by turning on $\lambda_{u}$ and $\lambda
_{d}$, driving the theory to a new infrared fixed point, the \textquotedblleft
IR theory\textquotedblright. More precisely, we assume that near the scale
$4\pi v$ with $v\simeq246$ GeV, the theory has flowed to an approximate
conformal phase, which can be viewed as the IR\ of the CFT. From the
perspective of electroweak symmetry breaking, however, this scale serves as a
UV cutoff. To keep the distinction clear, we will reference whether we are
discussing the IR\ of the CFT or the electroweak sector. We denote the scale
of conformal symmetry breaking as $M_{CFT}$. Below this scale, we can expect
the CFT to develop a mass gap. More precisely, we can expect a collection of
massive states as we pass from the CFT breaking scale down to the infrared of
the electroweak scale. These thresholds will occur at various mass scales, and
in general will depend on highly model dependent details of a given conformal
theory. Nevertheless, one of our aims will be to show that approximate
conformal symmetry still constrains the behavior of two-point functions in the
infrared of the electroweak theory. We shall also assume as in
\cite{HoloHiggs} that the mass spectrum is approximately supersymmetric, an
assumption which is well-justified in various mediation scenarios.

A number of properties of the new fixed point are specified by the infrared
R-symmetry of the mixed CFT. For example, the scaling dimension of any chiral
primary operator $\mathcal{O}$ is:%
\begin{equation}
\Delta_{IR}\left(  \mathcal{O}\right)  =\frac{3}{2}R_{IR}\left(
\mathcal{O}\right)  .
\end{equation}
To couple the visible sector to the CFT, we view the Standard Model gauge
group as a global symmetry of the CFT. The objects that show up in our
computation of $\Delta S$ and $\Delta T$ will then be two-point functions of
the currents that couple to these weakly gauged symmetries.

The infrared R-symmetry of the CFT\ also shows up in the calculation of
two-point functions of symmetry currents. Although we are interested in the
behavior of these two-point functions below the energy scale of conformal
symmetry breaking, it turns out that many of these properties can be recast in
terms of properties of the short-distance physics. This behavior is in turn
related to the operator product expansion (OPE) of the currents. In more
detail, recall that we need to calculate the behavior of the vacuum
polarization amplitude $\Pi_{ij}^{\mu\nu}(q^{2})$:%
\begin{equation}
\Pi_{ij}^{\mu\nu}(q^{2})=\left\langle J_{i}^{\mu}(q)J_{j}^{\nu}%
(-q)\right\rangle
\end{equation}
in the infrared of the electroweak theory. Here, $J_{i}^{\mu}(q)$ denotes the
(Fourier transform) of the flavor symmetry current, with flavor indices $i,j$
and Lorentz indices $\mu,\nu$. In a supersymmetric theory, this current sits
inside a supermultiplet $\mathcal{J}_{i}$, with
\begin{equation}
\mathcal{J}_{i}(x,\theta,\bar{\theta})=J_{i}(x)-\overline{\theta}\sigma_{\mu
}\theta J_{i}^{\mu}(x)+...
\end{equation}
See for example \cite{GGMI, GGMII, Dumitrescu:2011iu} for further discussion
of current supermultiplets. While the behavior of the correlation function in
the infrared of the electroweak theory involves many details about the mass
spectrum of the extra sector in the broken symmetry phase, the short distance
behavior in the CFT is tightly constrained by the operator product expansion
(OPE) for the symmetry currents. The OPE\ of the bosonic components $J_{i}$ is
(see e.g. \cite{Fortin:2011nq}):%
\begin{equation}
J_{i}(x)J_{j}(0)=b_{ij}\frac{1}{16\pi^{4}x^{4}}+d_{ijk}\frac{J_{k}(0)}%
{16\pi^{2}x^{2}}+w_{ij}\frac{K(0)}{4\pi^{2}x^{2-\gamma_{K}}}+c_{ij}^{l}%
\frac{\mathcal{O}_{l}(0)}{x^{4-\Delta_{i}}}+\text{descendants.}\label{JJOPE}%
\end{equation}
Here, $b_{ij}$ is an anomaly coefficient:%
\begin{equation}
b_{ij}=-3\text{Tr}(R_{IR}J_{i}J_{j}).\label{babs}%
\end{equation}
This trace is over the various species of Weyl fermions in the theory, and is
calculable via 't Hooft anomaly matching as long as we know the infrared
R-symmetry of the CFT. The terms with $i=j$ correspond to the numerator of the
NSVZ beta function. Contributions when $i\neq j$ can be viewed as including
the effects of kinetic mixing between different abelian symmetry factors. For
non-abelian symmetry currents such off-diagonal contributions vanish.

The remaining terms in equation (\ref{JJOPE}) include the operators
$\mathcal{O}_{i}(0)$ of dimension $\Delta_{i}$, which specify real conformal
primaries. $K(0)$ corresponds to the vev of the Konishi operator (i.e.
K\"{a}hler potential for weakly coupled fields), with anomalous dimension
$\gamma_{K}$. As discussed in \cite{Fortin:2011nq,Fortin:2011ad}, the OPE is
useful even if the flavor symmetries of the currents are spontaneously broken.

Clearly, a number of properties of visible/extra sector mixing are specified
by the IR R-symmetry of the new fixed point. Thankfully, this turns out to be
calculable in many superconformal theories of interest. In general, the IR
R-symmetry of the CFT is given by a linear combination of the UV R-symmetry
$R_{UV}$ with all infrared flavor symmetries $F_{i}$:%
\begin{equation}
R_{IR}=R_{UV}+\sum t_{i}F_{i}\label{IRtoUV}%
\end{equation}
with the coefficients $t_{i}$ fixed by a-maximization
\cite{Intriligator:2003jj}. An important remark is that the terms on the
right-hand side of (\ref{IRtoUV}) can always be grouped in terms of conserved
currents. Thus, another way to organize equation (\ref{IRtoUV}) is in terms of
a putative infrared R-symmetry, and all possible linear combinations of
conserved abelian flavor symmetries. The procedure of a-maximization
determines the coefficients $t_{i}$ so that the resulting R-symmetry is the
unique one inside the same supermultiplet as the stress tensor. As stated
above, finding this R-symmetry is equivalent to finding the scaling dimensions
of chiral primary fields. When the Higgs fields develop small anomalous
dimensions, we can treat the mixing between the Higgs and the extra sector as small.

\subsection{Symmetry Assumptions}

From the perspective of the extra sector, the weakly coupled gauge group of
the Standard Model is essentially a flavor symmetry. Throughout this work, we
shall assume that for an appropriate range of couplings and Higgs vevs, the
extra sector also enjoys an approximate custodial $SU(2)$ symmetry,
$SU(2)_{cust}$. To this end, we assume throughout that the extra sector has a
flavor symmetry $SU(2)_{W}\times SU(2)_{\widetilde{W}}\times U(1)_{\chi}$ with
$SU(2)_{cust}=SU(2)_{diag}\subset SU(2)_{W}\times SU(2)_{\widetilde{W}}$. In
this limit, the Higgs fields $H_{u}$ and $H_{d}$ assemble into a $2\times2$
matrix transforming as a bi-doublet $(\mathbf{2},\mathbf{2})$ of
$SU(2)_{W}\times SU(2)_{\tilde{W}}$, which decomposes as $\mathbf{3}%
\oplus\mathbf{1}$ under $SU(2)_{cust}$. The electroweak-breaking vacuum will
then preserve $SU(2)_{cust}$ when the vevs $v_{u}$ and $v_{d}$ are equal.

The Standard Model hypercharge and electric charge embed in $SU(2)_{W}\times
SU(2)_{\widetilde{W}}\times U(1)_{\chi}$ in the standard way:\footnote{Let us note
that in the literature, it is common to refer to these symmetries as
$SU(2)_{L}\times SU(2)_{R}\times U(1)_{\chi}$, even though a left- or
right-handed state could be charged under both non-abelian factors. To
minimize confusion, we shall avoid this terminology in what follows.}%
\begin{equation}
Y=T_{\widetilde{W}}+T_{\chi}\text{ \ \ and \ \ }Q=T_{W}+T_{\widetilde{W}%
}+T_{\chi}.
\end{equation}
where $T_{\widetilde{W}}$ and $T_{W}$ denote the $\sigma_{(3)}$ direction of
the respective $SU(2)$ factors. It will prove convenient to introduce two
additional linear combinations, which we refer to as the vector and axial
symmetry generators:%
\begin{equation}
T_{V}=T_{W}+T_{\widetilde{W}}\text{ \ \ and \ \ }T_{A}=T_{W}-T_{\widetilde{W}%
}.\label{axvec}%
\end{equation}
The generator $T_{V}$ embeds as the $\sigma_{(3)}$ direction of $SU(2)_{cust}$.
Let us emphasize that in our terminology the notion of \textquotedblleft
vector\textquotedblright\ and \textquotedblleft axial\textquotedblright%
\ generators only makes reference to the algebraic structure of the symmetry. In
a parity symmetric theory this is equivalent to introducing currents which act
on left- and right-handed states differently. Since we will be interested in
theories where a weakly coupled description may not be available, we shall
adhere to the conventions just outlined.

Even without introducing a mass scale into the CFT, we can break
electroweak symmetry by introducing the explicit $SU(2)_{W}$ breaking term
which preserves $U(1)_{A}$ and $U(1)_{V}$:
\begin{equation}
\mathcal{L}_{eff}\supset\int d^{2}\theta\text{ }(\lambda_{u}H_{u}%
^{(0)}\mathcal{O}_{u}^{(0)}+\lambda_{d}H_{d}^{(0)}\mathcal{O}_{d}%
^{(0)})+h.c.\label{mosdef}%
\end{equation}
so that only the electrically neutral components of the operators mix. This
leads to a distinct fixed point, but is often quite close to the original
fixed point if the Higgs dimension remains close to its free field value.

We shall also assume that there is a $%
%TCIMACRO{\U{2124} }%
%BeginExpansion
\mathbb{Z}
%EndExpansion
_{2}$ symmetry under which the axial current is odd and the vector current and
$U(1)_{\chi}$ current are even. Once we break electroweak symmetry, we can
identify this $\mathbb{Z}_{2}$ symmetry with the standard parity of a
vector-like supersymmetric theory. Since the two-point function $\Pi_{3Y}$ is
given by $\frac{1}{2}\left(  \Pi_{VY}+\Pi_{AY}\right)  $ while the two-point
function for $\Pi_{AY}$ is $-\frac{1}{2}\Pi_{AA}$, the full two-point function
$\Pi_{3Y}$ can be written as (see e.g. \cite{Peskin:1991sw}):%
\begin{equation}
\Pi_{3Y}=\frac{1}{4}\left(  \Pi_{VV}-\Pi_{AA}\right)  .\label{PI3YVVAA}%
\end{equation}
This presentation of the two-point function will prove quite helpful when we
study mass-independent contributions to the $S$-parameter.

\section{Wave Function Renormalization and the $T$-Parameter
\label{sec:TPARAM}}

Having discussed our general assumptions about conformal symmetry breaking and
flavor symmetries of the superconformal extra sector, we now compute the
contribution $\Delta T$ from the states in the extra sector. We follow the
same strategy outlined in \cite{DSSM}. Rather than directly evaluate the
current two-point functions, we shall instead consider the net effect the
extra sector has on the masses of the W- and Z-bosons, as reflected in the
$\rho$-parameter $\rho\equiv M_{W}^{2}/M_{Z}^{2}c_{W}^{2}$. This is in turn
related to $T$ via (see e.g. \cite{Peskin:1991sw}):
\begin{equation}
\rho=\frac{1}{1-\alpha_{EM}T}\simeq1+\alpha_{EM}T
\end{equation}
Hence, to compute $\Delta T$, it is enough to compute the change in the masses
of the gauge bosons from coupling to the extra sector.

To calculate this contribution, we view the Higgs vevs as a supersymmetric
modulus which introduces an effective mass into the superconformal extra
sector. The net response of the extra sector on the Higgs is then dictated by
a change in the effective K\"{a}hler metric on the moduli space of the Higgs
vevs. In a supersymmetric theory, these corrections show up in the
Coleman-Weinberg \cite{Coleman:1973jx} correction to the K\"{a}hler potential:%
\begin{equation}
\delta K=-\frac{1}{32\pi^{2}}\text{Tr}\left(  M^{\dag}M\log\frac{M^{\dag}%
M}{\mu_{(0)}^{2}}\right)  . \label{specialK}%
\end{equation}
where $M$ denotes the mass matrix of the theory, and $\mu_{(0)}$ denotes a UV
cutoff scale.

To estimate the corrections to the K\"{a}hler potential, we work to leading order
in conformal perturbation theory, that is, we restrict to contributions involving
two Higgs fields. We assume the existence of an approximate
custodial $SU(2)_{cust}$ symmetry. This means the
Higgs fields $H_{u}$ and $H_{d}$ can be assembled into a $2\times2$ matrix
$\Phi$ which transforms in the adjoint of $SU(2)_{cust}$, with a common
dimension:%
\begin{equation}
\Delta=1+\delta
\end{equation}
where $\delta$ denotes the excess Higgs dimension, once these fields mix with
the CFT. The two real flavor neutral bilinears we can form are $\det \Phi + h.c.$ and
$\Phi^{\dag} \Phi$. However, since terms entering our K\"{a}hler potential need to be invariant
under the infrared R-symmetry of the CFT, it follows that to leading order in conformal perturbation theory,
the K\"{a}hler potential is constrained to be $(\Phi^{\dag}\Phi)^{1/\Delta}$. In
more formal terms, $\Phi^{\dag} \Phi$ denotes the term of the OPE proportional
to the identity operator. Since $\delta$ is taken
to be small, we can take this operator to have scaling dimension
$\Delta_{\Phi^{\dag} \Phi} \simeq 2 \Delta_{\Phi}$.

Expanding out, we learn that the K\"{a}hler potential
for the Higgs fields is, in the approximate custodial limit:%
\begin{equation}
K=(\Phi^{\dag}\Phi)^{1/\Delta}=(H_{u}^{\dag}H_{u}+H_{d}^{\dag}H_{d}%
)^{1/\Delta}.
\end{equation}
This expression is independent of the Yukawas because we are working
in the special limit where the ratio $\lambda_{u}/\lambda_{d}\simeq
1$.\footnote{As a brief digression, it is of course tempting to extend this to
more general values of the Yukawas by rescaling the Higgs fields. Indeed, from
the perspective of the conformal sector, the only thing the
CFT\ \textquotedblleft knows about\textquotedblright\ is the overall mass
scale $m_{u}=\lambda_{u}v_{u}$ and $m_{d}=\lambda_{d}v_{d}$, so we can
assemble the corrections of equation (\ref{specialK}) into a single $2\times2$
matrix. However, what spoils the argument is that the overall coefficient of
this term relative to the original $H^{\dag}H$ terms is no longer constrained,
since they respect different custodial $SU(2)$ symmetries. Indeed, this is in
accord with the general expectation that away from the custodial $SU(2)$
limit, there will be a non-trivial dependence on the Yukawa couplings.}

This K\"{a}hler potential modifies the kinetic term for the Higgs doublet:%
\begin{equation}
\mathcal{L}_{kin}=-g_{i\overline{j}}\left(  D_{\mu}\Phi\right)  ^{i}\left(  D^{\mu}%
\Phi^{\dag}\right)  ^{\overline{j}}.
\end{equation}
As computed in \cite{DSSM}, the resulting change in the mass of the W- and
Z-bosons is then:%
\begin{equation}
M_{W}^{2}=\frac{g^{2}}{2\Delta}\left(  v_{u}^{2}+v_{d}^{2}\right)  ^{1/\Delta
}\text{ and }M_{Z}^{2}=\frac{M_{W}^{2}}{c_{W}^{2}}\times\frac{\left(
v_{u}^{2}+v_{d}^{2}\right)  ^{2}+4v_{u}^{2}v_{d}^{2}\left(  \Delta-1\right)
}{\Delta\left(  v_{u}^{2}+v_{d}^{2}\right)  ^{2}}.
\end{equation}
Expanding to leading order in $\delta$, we have:%
\begin{equation}
\rho=1+\left(  1-\frac{4v_{u}^{2}v_{d}^{2}}{\left(  v_{u}^{2}+v_{d}%
^{2}\right)  ^{2}}\right)  \delta=1+\cos^{2}2\beta\times\delta.
\end{equation}
Using the further relation $\rho=1/(1-\alpha_{EM}T)$, we obtain the estimate:%
\begin{equation}
\Delta T=\frac{1}{\alpha_{EM}}\cos^{2}2\beta\times\delta\label{DELTAT}%
\end{equation}
which vanishes in the custodial limit $\tan \beta \rightarrow 1$.
Using a-maximization, it is possible to compute $\delta$. In many cases
distinct from SQCD-like theories, $\delta$ can be small (see \cite{HVW} for
some examples). Summarizing, we have shown that in this
limit, $\Delta T$ is actually calculable, and is proportional
to the excess dimension of the Higgs.

\section{Anomalies and the $S$-Parameter \label{sec:SPARAM}}

We now turn to an evaluation of the $S$-parameter in scenarios where the Higgs
mixes with a conformal sector. From the perspective of the conformal theory,
this effect is more subtle, as it cannot be recast in terms of a wave-function
renormalization of the Higgs fields. Nevertheless, we will argue that enough
data is typically available in a conformal theory to estimate the leading
contributions to $\Delta S$ in terms of various calculable anomalies of the
conformal theory:%
\begin{equation}
\Delta S=\frac{b_{AA}}{4\pi}-\frac{4b_{W_{u}Y_{u}}}{2\pi}\log\frac{m_{u}%
}{m_{d}},
\end{equation}
with $b_{ij}$ an appropriate anomaly coefficient similar to that in equation
(\ref{babs}).

To organize the various contributions to the $S$-parameter, it is helpful to
review some general features of dispersion relations for two-point functions.
As explained in \cite{Peskin:1991sw}, this provides a way to relate the
$S$-parameter to the short distance behavior of the correlation function. By
definition, the contribution $\Delta S$ from states of the CFT comes
from evaluating $-16\pi\Pi_{3Y}^{\prime}(0)$, where we implicitly subtract off
the contribution from the Standard Model and supersymmetric 2HDM. We can
recast evaluation of $\Pi_{3Y}^{\prime}(0)$ as the residue integral of an
analytic function $\Pi_{3Y}^{\prime}(s)$ in the complex $s=q^{2}$ plane:
\begin{equation}
\Pi_{3Y}^{\prime}(0)=\frac{1}{2\pi i}%
%TCIMACRO{\doint }%
%BeginExpansion
{\displaystyle\oint}
%EndExpansion
\frac{ds}{s}\Pi_{3Y}^{\prime}(s).
\end{equation}
At a sufficiently large radius for the contour, we encounter
poles and branch cuts in $\Pi^{\prime}_{3 Y}(s)$, indicative
of the mass spectrum of the theory. For example, in the case
of a weakly coupled free field of mass $m$, there is a
pole at $s=m^{2}$ and a branch cut at $s=4m^{2}$, when the spectrum enters a
continuum. More generally, we can expect a complicated set of contributions,
depending on the details of the mass spectrum of the CFT after breaking
conformal symmetry.

Assuming suitable convergence properties of $\Pi_{3Y}^{\prime}(s)$, we can
take the residue integral to extend out to a circle of large radius. The
contour integral can then be replaced by contours encircling the poles in the
$s$-plane, and the branch cut. The main approximation we are going to make
throughout this work is that there is a branch cut which starts at
$s=M_{CFT}^{2}$, and that we can introduce a single contour to encircle all of
the poles. Contributions from the poles correspond to a given resonance, and
are manifestly finite contributions to $\Delta S$. Contributions from crossing
a branch cut lead to effects which depend logarithmically on the mass scales
of the theory. These branch cut effects are more subtle, as an individual
logarithm will superficially appear to involve some dependence on a cutoff. Of
course, the net contribution to $\Delta S$ must be cutoff independent.

Motivated by these general considerations, we shall parameterize the net
contribution to the $S$-parameter by grouping all of the poles into one
contribution, and the branch cut contributions into another:%
\begin{equation}
\Delta S=\Delta S_{(0)}-\frac{\delta b}{2\pi}\cdot\log f\left(  m_{u}%
,m_{d}\right)  \label{ourDeltaS}%
\end{equation}
where $f\left(  m_{u},m_{d}\right)  $ is a general function of $m_{u}$ and
$m_{d}$ which can be viewed as summing up the thresholds of masses in the
theory. We refer to $\Delta S_{(0)}$ as the mass-independent contribution,
since to leading order these details only appear in the logarithmic term. In
our parametrization, these extra terms vanish as $f\rightarrow1$ in the
custodial limit $m_{u}\rightarrow m_{d}$.

The rest of this section is organized as follows. First, we show that the
leading order contribution to $\Delta S_{(0)}$ is a calculable anomaly
coefficient. Next, we show that the leading contributions to the logarithmic
terms can also be calculated, provided the extra sector enjoys some additional
well-motivated symmetries. Finally, we also discuss the limit where the
Higgs vevs are a subdominant contribution to the mass of the extra sector. In
this limit we can also relate $\Delta S$ to the Higgs decay rates into
$\gamma\gamma$ and $Z\gamma$.

\subsection{The Mass-Independent Contribution}

In this section we turn to the calculation of the mass-independent
contribution $\Delta S_{(0)}$ in equation (\ref{ourDeltaS}). Rather than
working directly in terms of the value of the two-point function in the
infrared of the electroweak theory, we shall instead attempt to approximate
$\Delta S_{(0)}$ based on its high energy behavior.

The short-distance behavior of the current correlation functions is dictated
by the OPE reviewed in equation (\ref{JJOPE}):%
\begin{equation}
J_{i}(x)J_{j}(0)=b_{ij}\frac{1}{16\pi^{4}x^{4}}+d_{ijk}\frac{J_{k}(0)}%
{16\pi^{2}x^{2}}+w_{ij}\frac{K(0)}{4\pi^{2}x^{2-\gamma_{K}}}+c_{ij}^{l}%
\frac{\mathcal{O}_{l}(0)}{x^{4-\Delta_{l}}}+\text{descendants.}%
\end{equation}
The analogue of mass insertions in a weakly coupled model correspond to the
terms proportional to $c_{ij}^{l}\mathcal{O}_{l}(0)$, with $\mathcal{O}%
_{l}(0)$ real primaries of the superconformal theory. The operators which get
a non-zero vev are $H_{u}^{\dag}\sigma_{(l)}H_{u}$ and $H_{d}\sigma_{(l)}%
H_{d}^{\dag}$, which have dimension close to two when the Higgs/extra
sector mixing is small. This is in accord with the general discussion
near equation (\ref{OS}). What we are going to do is recast these insertions as
the first term in the OPE of an axial current two-point function.

Now, in the custodial $SU(2)$ limit, the Higgs field is a diagonal matrix
proportional to the identity, and thus leaves unbroken $U(1)_{V}\times
U(1)_{\chi}$. In other words, insertions of the Higgs fields will only
generate a contribution for the \textit{broken} symmetry generator $U(1)_{A}$.
This means that in the two-point function of equation (\ref{PI3YVVAA}):%
\begin{equation}
\Pi_{3Y}=\frac{1}{4}\left(  \Pi_{VV}-\Pi_{AA}\right)
\end{equation}
it is enough to track the behavior of $\Pi_{AA}$.

To approximate the behavior of $\Pi_{AA}$, we treat the effects of the massive
states as a single threshold correction, controlled by an overall beta function
coefficient $b_{AA}$. Matching the asymptotic behavior of the CFT expression to that
of the single threshold approximation, we will extract the mass-independent contribution
$\Delta S_{(0)}$. In other words, we approximate the
net contribution of the CFT to the two-point function as if it were a single
particle of mass $M$, with coupling to the gauge fields specified by its net
effect on the beta functions. To this end, it is helpful to recall the
behavior of a general threshold correction to a weakly gauged flavor symmetry.
This appears in the vacuum polarization amplitude:%
\begin{equation}
\Pi_{XX}^{\mu\nu}\left(  q^{2}\right)  =i\left(  \eta^{\mu\nu}q^{2}-q^{\mu
}q^{\nu}\right)  \widehat{\Pi}_{XX}\left(  q^{2},\mu_{(0)}^{2},M^{2}\right)
\end{equation}
In the regimes $\left\vert q\right\vert \gg M$ and $\left\vert q\right\vert
\ll M$, $\widehat{\Pi}_{XX}\left(  q^{2},\mu_{(0)}^{2},M^{2}\right)  $ behaves
as:%
\begin{equation}
\widehat{\Pi}_{XX}\left(  q^{2},\mu_{(0)},M^{2}\right)  =\frac{b_{XX}}%
{8\pi^{2}}\left\{
\begin{array}
[c]{l}%
\log\frac{M}{\mu_{(0)}}\text{ for }\left\vert q\right\vert \ll M\text{ }\\
\log\frac{M}{|q|} \,\,\, \text{ for }\left\vert q\right\vert \gg M
\end{array}
\right.  .\label{piprimethresh}%
\end{equation}
where $b_{XX}=-3\mathrm{Tr}(R_{IR}J_{X}J_{X})$ is the beta function
coefficient. The assumption that we can treat all of the states as a single
threshold means that the overall coefficient $b_{XX}$ tracks the mass-independent
contribution to $\Pi_{XX}^{\prime}(0)$.

Turning to our case, the single threshold approximation means that the mass-independent
contribution to the $S$-parameter is given by an anomaly coefficient
for the axial symmetry:%
\begin{equation}
\Delta S_{(0)}\simeq\frac{1}{2}\times-16\pi\times-\frac{1}{4}\times
\frac{b_{AA}}{8\pi^{2}}=\frac{b_{AA}}{4\pi}\label{massindep}%
\end{equation}
where:
\begin{equation}
b_{AA}=-3\text{Tr}(R_{IR}J_{A}J_{A}).\label{TRACE}%
\end{equation}
The overall factor of $1/2$ in equation (\ref{massindep}) is due to the
fact that in our $\mathbb{Z}_{2}$ symmetric theory, we must not double
count the contribution from a Dirac fermion.

It is interesting to note that in the models we are considering where the CFT
enjoys an $SU(2)_{W}\times SU(2)_{\widetilde{W}}\times U(1)_{\chi}$ symmetry,
the anomaly coefficient $b_{AA}$ is numerically the same as $b_{VV}$, the
contribution from the vector current. This is because after summing over all
components of a multiplet, the net contribution from $J_{W}J_{\widetilde{W}}$
vanishes. Note, however, that we can also extend our discussion to cover the
case detailed in equation (\ref{mosdef}), where we break by hand the
non-abelian symmetry. In this case, we can still compute $b_{AA}$, and now it
could differ from $b_{VV}$.

Finally, it is also instructive to compare our computation with the case of
technicolor-like theories, and other models where the Higgs boson is a
composite operator. This corresponds to allowing a possibly large correction
to the scaling dimension of the Higgs once it mixes with the SCFT. In such
cases, there is no sense in which there is a \textquotedblleft mass
insertion\textquotedblright\ by the Higgs fields. Indeed, this is reflected by
the fact that the vector and axial two-point functions both make sizable
contributions to the $S$-parameter, as for example in the estimate of
\cite{Peskin:1991sw} for QCD-like technicolor theories. Moreover, this leads
to a rather different behavior for the vacuum amplitude which is often
well-approximated by a \textquotedblleft vector-dominance\textquotedblright%
\ model of the scaled up QCD-like theory. The situation here is reversed, and
is more in line with the weakly coupled analysis. Here, the dominant
contribution is from the two-point function for the axial current. Observe,
however, that the behavior of the vector and axial two-point functions in the
UV\ regime of the CFT are the same, in accord with the fact that numerically,
$b_{AA}=b_{VV}$ in models with $SU(2)_{W}\times SU(2)_{\widetilde{W}}\times
U(1)_{\chi}$ symmetry.

\subsection{The Logarithmic Contribution}

In addition to the mass-independent contribution $\Delta S_{(0)}$, the general
behavior of the two-point function will depend on the mass spectrum of the
extra sector states. This is reflected in the mass threshold
function $f\left(  m_{u},m_{d}\right)  $ in equation (\ref{ourDeltaS}):
\begin{equation}
\Delta S=\Delta S_{(0)}-\frac{\delta b}{2\pi}\log f\left(  m_{u},m_{d}\right) .
\end{equation}
In general, we cannot hope to estimate this in full detail, as it involves
unrealistic expectations on the amount of information we have about the CFT.
However, in special cases, we can obtain an accurate approximation as follows.

We now show how to extract these contributions when the Hilbert space of
states splits into up-type and down-type states:%
\begin{equation}
\mathcal{H}=\mathcal{H}_{u}\oplus\mathcal{H}_{d}.
\end{equation}
More precisely, we consider the related system where we break by hand the
$SU(2)_{W}$ symmetry by omitting the fields $H_{u}^{(+)}$ and $H_{d}^{(-)}$
from the visible sector. In this limit, the flavor symmetry considered
previously is $U(1)_{W}\times U(1)_{\widetilde{W}}\times U(1)_{\chi}$. The
up-down approximation amounts to the further assumption that the flavor
symmetry contains two copies of this symmetry, which we denote by
$G_{up}\times G_{down}$. We will denote the up-type symmetries by
$U(1)_{W_{u}}\times U(1)_{\widetilde{W_{u}}}\times U(1)_{\chi_{u}}$, and the
down-type symmetries by $U(1)_{W_{d}}\times U(1)_{\widetilde{W_{d}}}\times
U(1)_{\chi_{d}}$. The two hypercharges are given by $Y_{u,d} = T_{\widetilde
W_{u,d}} + T_{\chi_{u,d}}$.

Although this might seem like an ad hoc assumption, it is well-motivated from
both bottom-up and top-down considerations. For example, in a generic model
with two Higgs doublets, there can often be problems with flavor-changing
neutral currents. Such additional symmetries can naturally suppress
potentially dangerous contributions to flavor-changing neutral currents. This
is the case in the MSSM, where couplings between the up-type states and
$H_{d}$ do not appear holomorphically. Additionally, from a top-down
perspective, such a splitting occurs in many string/F-theory
constructions, a feature we will study more in section \ref{sec:EXAMPLES}.

Combining this with our previous discussion near equation (\ref{piprimethresh}%
), we deduce the further contribution to the two point function:
\begin{equation}
\Delta S-\Delta S_{(0)}=-\frac{b_{W_{u}Y_{u}}}{\pi}\log\frac{m_{u}^{2}}%
{\mu_{(0)}^{2}}-\frac{b_{W_{d}Y_{d}}}{\pi}\log\frac{m_{d}^{2}}{\mu_{(0)}^{2}}%
\end{equation}
where the relevant anomaly coefficients are:%
\begin{align}
b_{W_{u}Y_{u}} &  =-3\text{Tr}(R_{IR}J_{W_{u}}J_{Y_{u}})\\
b_{W_{d}Y_{d}} &  =-3\text{Tr}(R_{IR}J_{W_{d}}J_{Y_{d}}).
\end{align}
Since in the fully theory the states of the CFT fill out complete
$SU(2)_{W}\times SU(2)_{\widetilde{W}}$ multiplets, we also have $b_{W_{u}Y_{u}%
}=-b_{W_{d}Y_{d}}$. In other words, the full estimate for the $S$-parameter is:%
\begin{equation}
\Delta S=\frac{b_{AA}}{4\pi}-\frac{4b_{W_{u}Y_{u}}}{2\pi}\log\frac{m_{u}%
}{m_{d}}%
\end{equation}
and all dependence on the UV\ cutoff $\mu_{(0)}$ has dropped out, as it must. The
factor of $-4$ is due to the factor of $-1/4$ appearing in equation
(\ref{PI3YVVAA}). Since this contribution is not $%
%TCIMACRO{\U{2124} }%
%BeginExpansion
\mathbb{Z}
%EndExpansion
_{2}$ symmetric, there is also an additional factor of $2$ relative to the
mass-independent contribution, as per our discussion near equation
(\ref{massindep}).

In other words, we have reduced the computation of the $S$-parameter to the
computation of two anomaly coefficients. This is clearly a significant
simplification, and can often be extracted without detailed knowledge of the CFT.

\subsection{Vector-Like Limit and Higgs Decays \label{ssec:vec}}

The vector-like limit amounts to switching on large mass terms in the extra sector
which preserve the symmetries $SU(3)_{C}\times SU(2)_{W}\times U(1)_{Y}$.
For a characteristic mass scale $\Lambda_{vec}$, the contribution to the
$S$-parameter decouples as $v^{2}/\Lambda_{vec}^{2}$. In
this subsection we first specify the dimension six supersymmetric operator
related to the $S$-parameter and then show that this is also related to dimension
five operators which control Higgs decays to $\gamma\gamma$ and $Z\gamma$.
These decays are particularly interesting as probes of new physics as they are
only generated radiatively.

When supersymmetry is preserved, the kinetic terms for the gauge fields are:%
\begin{equation}
\mathcal{L}_{eff}=\operatorname{Im}\int d^{2}\theta\text{ }\frac{\tau_{W} }{8\pi}\text{Tr}_{SU(2)}\mathcal{W}^{\alpha}\mathcal{W}_{\alpha
}+\operatorname{Im}\int d^{2}\theta\text{ }\frac{\tau_{Y}}{8\pi}\mathcal{B}^{\alpha}\mathcal{B}_{\alpha}.
\end{equation}
with $\mathcal{W}_{\alpha}$ and $\mathcal{B}_{\alpha}$ the superfield
strengths for $SU(2)_{W}$ and $U(1)_{Y}$, respectively. Here, the holomorphic
gauge coupling is:%
\begin{equation}
\tau_{(i)}=\frac{4\pi i}{g_{(i)}^{2}} + \frac{\theta_{(i)}}{2\pi}.
\end{equation}
Integrating out the states of the extra sector, we read
off $\Delta S$ from the single holomorphic dimension six operator
which mixes $SU(2)_{W}$ and $U(1)_{Y}$:%
\begin{equation}
\mathcal{O}_{HHWB}=\operatorname{Re}\int d^{2}\theta\text{ \ }\frac{-b_{mix}}{8\pi}%
\cdot\frac{H_{u}\sigma^{(i)}H_{d}\mathcal{W}_{(i)}^{\alpha}\mathcal{B}%
_{\alpha}}{\Lambda_{mix}^{2}}.
\end{equation}
The energy scale $\Lambda_{mix}$ is roughly related to the mass of the extra
states via $\lambda_{u}\lambda_{d}\Lambda_{mix}^{2}\simeq M^{2}$.\footnote{There
can be further suppression if there is an approximate Peccei-Quinn symmetry
in the Higgs sector.}

Comparing with our discussion near line (\ref{OS}), we have $b_{mix}%
\simeq16\pi c_{S}^{(mix)}$. This specifies a different limit from the one
where the Higgs is the sole source of mass. Indeed, with a conserved
$SU(2)_{W}\times U(1)_{Y}$ flavor symmetry and approximate supersymmetry in
the extra sector, the coefficients $c_{u}$ and $c_{d}$ in equation (\ref{OS})
are suppressed by a factor of $\mu_{MSSM}/\Lambda_{mix}$ relative to
$c_{S}^{(mix)}$. Switching on the Higgs vevs $\left\langle H_{u}%
^{0}\right\rangle =v_{u}/\sqrt{2}$ and $\left\langle H_{d}^{0}\right\rangle
=v_{d}/\sqrt{2}$, we obtain:%
\begin{equation}
\Delta S=\frac{4s_{W}c_{W}}{\alpha_{EM}}\frac{b_{mix}}{16\pi}\frac{v_{u}v_{d}%
}{\Lambda_{mix}^{2}}.
\end{equation}
In a CP-preserving supersymmetric two Higgs doublet model, $v_{u}v_{d}%
/\Lambda_{mix}^{2}>0$. So in other words, the sign of $b_{mix}$ fixes the sign
of $\Delta S$.

The coefficient $b_{mix}$ is also closely connected with decays of the Higgs
into $\gamma\gamma$ and $Z\gamma$. The other dimension six operators
compatible with gauge invariance and holomorphy are:%
\begin{align}
\mathcal{O}_{HHWW}  &  =\operatorname{Re}\int d^{2}\theta\text{ \ }\frac{-b_{W}}{8\pi
}\cdot\frac{H_{u}H_{d}\text{Tr}_{SU(2)_{W}}\mathcal{W}^{\alpha}\mathcal{W}%
_{\alpha}}{\Lambda_{W}^{2}}\\
\mathcal{O}_{HHYY}  &  =\operatorname{Re}\int d^{2}\theta\text{ \ }\frac{-b_{Y}}{8\pi
}\cdot\frac{H_{u}H_{d}\mathcal{B}^{\alpha}\mathcal{B}_{\alpha}}{\Lambda
_{Y}^{2}}\\
\mathcal{O}_{HH\gamma\gamma}  &  =\operatorname{Re}\int d^{2}\theta\text{ \ }%
\frac{-b_{EM}}{16\pi}\cdot\frac{H_{u}H_{d}\mathcal{E}^{\alpha}\mathcal{E}%
_{\alpha}}{\Lambda_{EM}^{2}}. \label{HEE}%
\end{align}
with $\mathcal{E}_{\alpha}$ the supersymmetric field strength for $U(1)_{EM}$,
and $b_{G}=-3$Tr$(R_{IR}J_{G}J_{G})$ calculable beta function
coefficients. Assuming that there is only weak Higgs/SCFT mixing, we can
ignore the distinction between the physical and holomorphic gauge couplings.
Expanding the operator $\mathcal{O}_{HH\gamma\gamma}$ in the 2HDM mass eigenstate basis
yields the mixing angle dependence and sign of possible contributions to
$h^{0}\rightarrow\gamma\gamma$, $H^{0}\rightarrow\gamma\gamma$ and
$A^{0}\rightarrow\gamma\gamma$ \cite{HoloHiggs}. A similar analysis also
applies to decays to $Z\gamma$ by adding up the contributions from $\mathcal{O}_{HHWB}$,
$\mathcal{O}_{HHWW}$ and $\mathcal{O}_{HHYY}$. This yields the same 2HDM\ mixing angle dependence
found in \cite{HoloHiggs}.

Finally, switching on the Higgs vevs for the dimension six
operators and matching the contribution of $\mathcal{O}_{HHWB}+\mathcal{O}_{HHWW}+\mathcal{O}_{HHYY}$ to
that of $\mathcal{O}_{HH\gamma\gamma}$, we also learn that $\Delta S$ is related to the
beta function coefficients:%
\begin{equation}
\frac{b_{mix}}{\Lambda_{mix}^{2}}=\frac{b_{EM}}{\Lambda_{EM}^{2}}-\frac{b_{Y}%
}{\Lambda_{Y}^{2}}-\frac{b_{W}}{\Lambda_{W}^{2}}\text{.}%
\end{equation}

\section{Unitarity Constraints \label{sec:UNIT}}

Having recast the contributions to $\Delta S$ and $\Delta T$ in terms of
calculable anomaly coefficients, we now extract model-independent bounds
present in any unitary conformal theory. This enables us to impose constraints
on the relative size, and sign of these contributions.

Consider first the contribution to $T$, given by equation (\ref{DELTAT}):%
\begin{equation}
\Delta T=\frac{1}{\alpha_{EM}}\cos^{2}2\beta\times\delta.
\end{equation}
Even if the Higgs dimension deviates from its free field value, $\Delta T$ is small
when $\tan \beta \simeq 1$. The important feature for us is that $\cos^{2}2\beta$ is a number between zero
and one which is always non-negative. In principle, $\delta$ could be either
positive or negative. However, in a unitary conformal theory where the Higgs
is a chiral primary, the sign is fixed so that $\delta\geq0$ (see e.g.
\cite{Mack:1975je, Grinstein:2008qk}). To get a negative sign for the excess
Higgs dimension, one would need to either be in a non-conformal phase, or in a
limit where the Higgs is not a gauge-invariant operator. This could happen if
the Standard Model gauge group becomes strongly coupled, though this is rather
antithetical to the set of assumptions made in this work. We therefore
conclude that in most motivated situations, $\Delta T\geq0$. This is also in
accord with the weakly coupled expression in equation (\ref{Tweak}) near the
custodial $SU(2)$ limit. Expanding $m_{1}=m_{2}(1+\varepsilon)$, one obtains a
positive definite contribution to $\Delta T$ in this case as well.

Consider next the contribution to $S$:%
\begin{equation}
\Delta S=\frac{b_{AA}}{4\pi}-\frac{4b_{W_{u}Y_{u}}}{2\pi}\log\frac{m_{u}%
}{m_{d}}%
\end{equation}
which depends on the beta function coefficients $b_{AA}$ and $b_{W_{u}Y_{u}}$.
Even without knowing the precise value of these coefficients, we can still
deduce some valuable information. First of all, we have that in a general
unitary theory, the matrix $b_{ij}$ has non-negative eigenvalues, that is, it
is non-negative definite \cite{Anselmi:1997am, Anselmi:1997ys}. In the present
case with flavor symmetry $U(1)_{W}\times U(1)_{\widetilde{W}}\times
U(1)_{\chi}$, the $3\times3$ matrix of beta function coefficients:%
\begin{equation}
b_{ij}=\left[
\begin{array}
[c]{ccc}%
b_{VV} &  & \\
& b_{AA} & \\
&  & b_{\chi\chi}%
\end{array}
\right]  \label{3by3}%
\end{equation}
is non-negative definite. All off-diagonal elements vanish due to our
$\mathbb{Z}_{2}~$symmetry, and the fact that the vector and axial currents
embed in non-abelian symmetries (and so do not mix with $U(1)_{\chi}$). From
this, we deduce that $b_{AA}\geq0$. Putting this together, we learn that when
$m_{u}=m_{d}$, the contribution to $S$ is fixed to be positive. A simple way
to evade this bound is to consider a non-conformal extra sector. This is
related to the fact that spin $1$ particles contribute to this two-point
function with the opposite sign of spin $1/2$ or $0$ particles.

Consider next the logarithmic contributions to $\Delta S$. In contrast to the
structure of equation (\ref{3by3}), here, there can in principle be mixing
terms between $U(1)_{V}$ and $U(1)_{\chi}$, because the non-abelian structure
is no longer present, as we have projected down to just the up-type states. In
terms of the vector and axial two point functions, we have:%
\begin{equation}
b_{W_{u}Y_{u}}=\frac{1}{4}\left(  b_{V_{u}V_{u}} + 2 b_{V_{u}\chi_{u}}%
-b_{A_{u}A_{u}}\right)  .
\end{equation}
which could have either sign. Unitarity imposes the milder constraints:%
\begin{equation}
b_{V_{u}V_{u}}\cdot b_{\chi_{u}\chi_{u}}\geq \frac{1}{4}\left(  b_{V_{u}\chi_{u}}\right)
^{2}, \,\,\,\, b_{V_{u} V_{u}} + b_{\chi_{u}\chi_{u}} \geq 0, \,\,\,\, b_{A_{u}A_{u}}\geq0\text{.}%
\end{equation}

Putting these considerations together, we conclude that while the mass-independent
contribution to $\Delta S$ is non-negative definite, there can be
deviations away from this value, corresponding to the corrections
proportional to $\log(m_{u}/m_{d})$. These other contributions can in
principle have either sign, providing a simple way to produce a smaller net
value of $\Delta S$. Of course, one must be careful about over-interpreting
this result. If we work in a regime where the logarithmic correction starts to
dominate over the constant shift from $b_{AA}$, the threshold approximation
adopted here is on the verge of breaking down. Nevertheless, it does indicate
that in principle, either sign of $\Delta S$ is possible.

\section{Examples \label{sec:EXAMPLES}}

In this section we discuss some examples, showing how to compute the leading
order contributions to $S$ and $T$ from an extra sector. Even at weak
coupling, our parametrization of $S$ is helpful, as it provides a simple way
to capture the leading order contributions from wave function renormalization.
First, we consider a weakly coupled example to fix our conventions. After
this, we discuss the analogue of this result when the extra sector involves
SQCD-like dynamics. We then turn to some top-down examples motivated by string constructions.

Let us emphasize that our aim here is not to construct realistic models, but
rather, to illustrate the utility of our computational framework. Indeed,
contributions from the two Higgs doublet sector can, for appropriate mixing
angles, counteract shifts from the extra sector (see e.g. \cite{DSSM,
Bellantoni:2012ag}). A more complete study would also involve a combined
analysis with present constraints on Higgs production and decays, which can
also be computed \cite{HoloHiggs}. It would be interesting to perform this
more complete analysis.

\subsection{Weakly Coupled Models}

We first consider a weakly coupled example. In this case, the calculation of
$T$ is essentially the same as in the weakly coupled setting, so we restrict
our discussion to $\Delta S$. Our aim will be to show that our general
analysis reproduces the expected weakly coupled result. To fix conventions, we
consider a collection of chiral superfields $N_{L,R}$ and $E_{L,R}$ with
quantum numbers:%
\begin{equation}%
\begin{tabular}
[c]{|c|c|c|c|}\hline
& $U(1)_{W}$ & $U(1)_{\widetilde{W}}$ & $U(1)_{\chi}$\\\hline
$N_{L}$ & $+1/2$ & $0$ & $+\chi$\\\hline
$E_{L}$ & $-1/2$ & $0$ & $+\chi$\\\hline
$N_{R}$ & $0$ & $-1/2$ & $-\chi$\\\hline
$E_{R}$ & $0$ & $+1/2$ & $-\chi$\\\hline
\end{tabular}
\ \ \ .
\end{equation}
We denote by $Y$ the hypercharge of the \textquotedblleft left-handed
doublet\textquotedblright. In this model, $Y=\chi$. Note that here, the
\textquotedblleft right-handed\textquotedblright\ states are actually
specified by chiral superfields, just as the left-handed states. This is
always possible to do, by complex conjugation of a given field.

This toy model possesses the up-type and down-type splitting we want; we take the up-type
fields to be $N_{L,R}$ and down-type fields to be $E_{L,R}$. The mass terms of
the theory are:%
\begin{equation}
W_{mass}=m_{d}E_{L}E_{R}+m_{u}N_{L}N_{R}%
\end{equation}
which are generated by the Higgs vevs. In the case of a weakly coupled
doublet, we can also set $\chi=-1/2$, corresponding to an interaction term in
the unbroken phase:%
\begin{equation}
W_{int}=\lambda_{d}H_{d}LE_{R}+\lambda_{u}H_{u}LN_{R}.\label{Wint}%
\end{equation}
The operators of the extra sector which communicate with the
Higgs fields are $\mathcal{O}_{u}=LN_{R}$ and $\mathcal{O}_{d}=LE_{R}$.

To compute the contributions to $S$, we need to evaluate the anomaly
coefficients $b_{AA}$ and $b_{W_{u}Y_{u}}$. The R-charge of a free Weyl fermion is $-1/3$, so the
axial contribution is:%
\begin{equation}
b_{AA}=-3\text{Tr}\left(  R_{IR}J_{A}J_{A}\right)  =-3\times\frac{-1}{3}%
\times\left(  4\cdot\left(  \frac{1}{2}\right)  ^{2}\right)  =1.
\end{equation}
We can also evaluate the contribution to $b_{W_{u}Y_{u}}$ from just the
up-type states. In this case, we observe that the only up-type state charged
under $SU(2)_{L}$ is $N_{L}$. Hence, we have:%
\begin{equation}
b_{W_{u}Y_{u}}=-3\text{Tr}\left(  R_{IR}J_{W_{u}}J_{Y_{u}}\right)
=-3\times\frac{-1}{3}\times\left(  \left(  \frac{1}{2}\right)  \cdot
\chi\right) = \frac{Y}{2}.
\end{equation}
The overall contribution to $S$ is therefore:
\begin{equation}
\Delta S=\frac{1}{2\pi}\left(  \frac{b_{AA}}{2}-4b_{W_{u}Y_{u}}\log\frac
{m_{u}}{m_{d}}\right)  =\frac{1}{2\pi}\left(  \frac{1}{2}-2Y\log\frac{m_{u}%
}{m_{d}}\right)  .
\end{equation}
Compared to the weakly coupled formulae of equations (\ref{Sweak}) and (\ref{weakS})
reviewed in section \ref{sec:REVIEW}, the overall contribution is larger
by a factor of $3/2$. This is because each chiral multiplet
contains a Weyl fermion, and a complex scalar.

Part of the point of our formalism is that it provides a simple way to
incorporate the effects of wave function renormalization for states of the
extra sector. These effects are encoded in the anomalous dimension $\gamma$
of a field, which is related to the total dimension as:%
\begin{equation}
\Delta=1+\frac{\gamma}{2}.
\end{equation}
Now, in a chiral multiplet, the R-charge of the Weyl fermion is $R-1$, where
$R$ is the R-charge of the complex scalar. This is in turn related to the
anomalous dimension via:%
\begin{equation}
R-1=\frac{2}{3}\Delta-1=-\frac{1-\gamma}{3}.
\end{equation}
The resulting contribution to $S$ is given by plugging in the R-charge
assignment for the Weyl fermions of the chiral multiplet:%
\begin{equation}
\Delta S=\frac{1-\gamma}{2\pi}\left(  \frac{1}{2}-2Y\log\frac{m_{u}}{m_{d}%
}\right)  .
\end{equation}
We see that as the anomalous dimensions of the extra sector fields increase,
$\Delta S$ decreases. While it might therefore seem possible to alter the sign
of $\Delta S$ by increasing $\gamma$, the amount it can increase is bounded
above. This is because in a unitary superconformal theory $b_{AA}\geq0$. We
will see an example of such a bound when we turn to SQCD-like extra sectors.
Sharper upper bounds for scaling dimensions in SCFTs have been studied
in \cite{Hook:2012fp, Amariti:2012wc}. In the case where we take into
account wave-function renormalization effects, we can also calculate
the leading order shift to $T$:%
\begin{equation}
\Delta T=\frac{1}{\alpha_{EM}}\cos^{2}2\beta\times\frac{\gamma}{2}.
\label{tagain}%
\end{equation}

It is also of interest to ask how much tuning is required in this toy model to keep the
contributions $\Delta S$ and $\Delta T$ smaller than an order $0.1$ correction to the Standard
Model reference value. Let us consider the special case $\lambda_u = \lambda_d$. In
the idealized situation where $\Delta S = 0$, we can solve for $\tan \beta$ in terms of $Y$:
\begin{equation}
\tan \beta = \exp\left(\frac{1}{4Y} \right).
\end{equation}
The value of $\Delta T$ is then:
\begin{equation}
\Delta T\simeq\frac{1}{\alpha_{EM}}\tanh^{2}\left(  \frac{1}{4Y}\right)
\times\frac{\gamma}{2}%
\end{equation}
which can in principle be small. Of course, it is a matter of taste (or
lack thereof) to decide how much tuning to accept in such situations.

A similar analysis applies in cases where the corresponding chiral multiplet
is charged under both $SU(2)_{W}$ and $SU(2)_{\widetilde{W}}$, as considered
near equation (\ref{weakS}):
\begin{equation}%
\begin{tabular}
[c]{|c|c|c|c|}\hline
& $U(1)_{W}$ & $U(1)_{\widetilde{W}}$ & $U(1)_{\chi}$\\\hline
$\psi_{L}$ & $+t_{L}$ & $+t_{R}$ & $+\chi$\\\hline
$\psi_{R}^{c}$ & $-t_{R}$ & $-t_{L}$ & $-\chi$\\\hline
$m_{(\psi)}$ & $-(t_{L}-t_{R})$ & $-(t_{R}-t_{L})$ & $0$\\\hline
\end{tabular}
\end{equation}
Using our general formulae, we reproduce equation (\ref{weakS}). Note that
here we have adopted a convention in which all \textquotedblleft
right-handed\textquotedblright\ states have been conjugated to left handed
states. Additionally, we have only written one component of a given
$SU(2)_{W}\times SU(2)_{\widetilde{W}}$ multiplet. Observe that once we sum
over a complete multiplet, we obtain $b_{AA}=b_{VV}$ in the conformal phase.

\subsection{SQCD-like Sectors}

We now turn to a more involved example where the states of the extra
sector participate in an SQCD-like theory. To this end, consider a theory
with gauge quantum numbers:%
\begin{equation}%
\begin{tabular}
[c]{|c|c|c|c|c|}\hline
& $U(1)_{W}$ & $U(1)_{\widetilde{W}}$ & $U(1)_{\chi}$ & $SU(N_{c})_{extra}%
$\\\hline
$N_{L}$ & $+1/2$ & $0$ & $+\chi$ & $N_{c}$\\\hline
$E_{L}$ & $-1/2$ & $0$ & $+\chi$ & $N_{c}$\\\hline
$N_{R}$ & $0$ & $-1/2$ & $-\chi$ & $\overline{N_{c}}$\\\hline
$E_{R}$ & $0$ & $+1/2$ & $-\chi$ & $\overline{N_{c}}$\\\hline
\end{tabular}
\ \ \ .
\end{equation}
where we take $g$ generations of each chiral superfield. In this case, the
mixing between the superconformal sector and the Higgs is given by the
analogue of equation (\ref{Wint}). This interaction term now drives the theory
to a new fixed point, mixing the Higgs and conformal sector. Weak mixing
corresponds to a short flow to the new fixed point, where the Higgs field
retains a scaling dimension close to its weakly coupled value.

First consider the behavior of the conformal theory in the absence of Higgs/extra
sector mixing. The extra sector is then a supersymmetric version of QCD. For an appropriate
number of flavors, this theory will be in the conformal window. The symmetries of this toy model imply
that the chiral fields of the extra sector have the same R-charge, which we denote by $R$.
The value of $R$ is fixed, since vanishing of the $SU(N_{c})_{extra}$ beta function yields the
condition:%
\begin{equation}
N_{c}+\frac{4g}{2}\left(  R-1\right)  =0
\end{equation}
or:%
\begin{equation}
R=1-\frac{N_{c}}{2g}\text{.}\label{rcharge}%
\end{equation}

Once we introduce the coupling to the Higgs fields, the R-charge assignments
in the infrared will consequently adjust. Since we are working in the limit
where there is an approximate custodial $SU(2)$ symmetry, the fields $H_{u}$
and $H_{d}$ have the same R-charge, which we denote by $R_{H}$. Moreover,
because the beta function condition is the same, the R-charge assignment of
equation (\ref{rcharge}) is the same. Imposing the further condition that the
superpotential has R-charge $2$ in the infrared, we also have the condition:%
\begin{equation}
R_{H}=\frac{N_{c}}{g}.
\end{equation}

Let us now turn to the calculation of $\Delta S$ and $\Delta T$. First of all,
the excess Higgs dimension is:%
\begin{equation}
\delta_{H}=\Delta_{H}-1=\frac{3N_{c}-2g}{2g}.
\end{equation}
The regime of weak Higgs/SCFT mixing corresponds to $2 g \simeq 3 N_{c}$, which is near the top
of the conformal window. Using equation (\ref{DELTAT}), we therefore have:%
\begin{equation}
\Delta T=\frac{1}{\alpha_{EM}}\cos^{2}2\beta\times\left(  \frac{3N_{c}-2g}%
{2g}\right)  .
\end{equation}
Consider next $\Delta S$. We first evaluate the anomaly coefficient $b_{AA}$:%
\begin{equation}
b_{AA}=-3\text{Tr}\left(  R_{IR}J_{A}J_{A}\right)  =-3\left(  R-1\right)
\times\left(  4 g N_{c} \cdot\left(  \frac{1}{2}\right)  ^{2}\right)  .
\end{equation}
Using equation (\ref{rcharge}), we learn that the anomaly is:%
\begin{equation}
b_{AA}=\frac{3}{2}N^{2}_{c}.
\end{equation}
The other anomaly we need to evaluate is $b_{W_{u}Y_{u}}$:
\begin{equation}
b_{W_{u}Y_{u}}=-3\text{Tr}\left(  R_{IR}J_{W_{u}}J_{Y_{u}}\right)
=-3\times\left(  R-1\right)  \times\left(  g N_{c} \left(  \frac{1}{2}\right)
\cdot\chi\right)
\end{equation}
from which we obtain:%
\begin{equation}
b_{W_{u}Y_{u}}=\frac{3\chi}{4}N^{2}_{c}.
\end{equation}
Thus, the overall contribution to $S$ is:%
\begin{equation}
\Delta S=\frac{3N^{2}_{c}}{4\pi}\left(  \frac{1}{2}-2Y\log\frac{m_{u}}{m_{d}%
}\right)
\end{equation}
where we have used $Y = \chi$ in this model.

\subsection{F-theory Inspired Scenarios}

We now turn to some examples motivated from F-theory model building
(for reviews see \cite{HVLHC, Heckman:2010bq, Weigand:2010wm, Maharana:2012tu}). In
F-theory models, the visible sector is obtained from intersecting
seven-branes, while probe D3-branes can serve as superconformal extra sectors
\cite{Funparticles}. Visible sector matter fields descend
from six-dimensional fields which propagate in our four spacetime dimensions
as well as two compact dimensions of the internal geometry. When the D3-brane
is close to an E-type Yukawa point, this leads to a strongly coupled
$\mathcal{N}=1$ superconformal theory. The flavor symmetry of the theory
contains the Standard Model gauge group, embedded in $E_{8}\supset
SU(5)_{GUT}\times SU(5)_{\bot}$. These theories can be viewed as
$\mathcal{N}=1$ deformations of the $\mathcal{N}=2$ supersymmetric
Minahan-Nemeschansky theory \cite{MNI, MNII}. Moreover, although they
are~$\mathcal{N}=1$ theories, they share many features of their $\mathcal{N}%
=2$ precursors, allowing for a quantitative study of some aspects of the
theory \cite{FCFT, HVW, D3gen}.

In contrast to weakly coupled models, the way that the states of the D3-branes pick up a
mass is more subtle because strong coupling effects play a role in generating mass terms. Even so,
moving the D3-brane away from the E-type point can be modelled in terms of $(\mathbf{5}\oplus\overline
{\mathbf{5}})^{\prime}s$ which are coupled to a strong $U(1)_{D3}$
\cite{Funparticles}. The effective number of these $(\mathbf{5}\oplus
\overline{\mathbf{5}})^{\prime}s$ is irrational, reflecting the fact that the
anomalous dimensions of these states are non-trivial. By moving along the direction
of one of the Higgs curves (say, the $H_{u}$), we can see that activating a
vev for the $H_{u}$ leads to a pairing between the electroweak
doublets of the $\overline{\mathbf{5}}$ and
corresponding singlets of the extra sector. In this limit, the colored
triplets of the $\overline{\mathbf{5}}$ do not pick up a mass, but since they
are neutral under $SU(2)_{W}$, they do not enter into the electroweak
parameters anyway. Combining this observation with anomaly matching considerations, we can
compute $\Delta S$:
\begin{equation}
\Delta S=\frac{N_{eff}}{2\pi}\left(  \frac{1}{2}+\log\frac{m_{u}}{m_{d}%
}\right)
\end{equation}
Here, the effective number $N_{eff}=\delta b_{SU(5)}$ is the contribution to
the $SU(5)$ beta function from the extra sector states. There can be
additional shifts to this number from coupling to the Higgs fields (which
explicitly breaks $SU(5)_{GUT}$) but these effects turn out to be subleading
\cite{HVW}. In typical F-theory examples, $\delta b_{SU(5)}\simeq2-4$ (see
e.g. \cite{HVW}). The contribution to $\Delta T$ involves knowing $\delta$,
the excess dimension of the Higgs. In typical examples, this can be on the order
of $0.01$ to $0.1$ \cite{HVW}.

\section{Conclusions \label{sec:CONC}}

The detection of the Higgs boson opens a doorway to studying signatures of
physics beyond the Standard Model. One well-motivated class of such scenarios
involves mixing the Higgs with an extra sector, which in many cases of
interest can be superconformal. In this work we have shown that it is possible
to extract the leading-order contributions to $\Delta S$ and $\Delta T$ from
calculable data of the extra sector such as anomaly coefficients. We have also
provided model-independent bounds on the size and sign of possible
contributions to $\Delta S$ and $\Delta T$, and illustrated in some examples
the applicability of these results. In the remainder of this section we
discuss potential directions of future investigation.

Since the coupling to the extra sector can affect the Higgs mass, a natural
next step would be to calculate these contributions in conformal perturbation theory.
It would be interesting to see how our estimates for various two-point functions feed into such
calculations, and it may be possible to use the information we already have to
compute these mass effects, perhaps along the lines of \cite{DSSM}.

One can view our analysis as part of a more general program to characterize
the possible higher-dimension operators involving Higgs fields. In previous
work \cite{HoloHiggs} an analysis in terms of anomaly coefficients was
used to deduce the leading order contributions to gluon fusion and diphoton
decays for Higgs/SCFT mixing. Combining this with our present analysis
provides a way to impose further constraints, and to discover possibly favored
corners of parameter space. A detailed numerical scan of such scenarios would
no doubt be useful.

Finally, in addition to its potential relevance for phenomenological analyses,
the methods used in this paper are also of interest from a purely formal
standpoint. More broadly, one can introduce the analogue of the $S$ and $T$
parameters for a general superconformal theory. Note that since the leading
order behavior of the correlation functions are calculable and rely only on
information about global symmetries, it is possible to use our methods to
study SCFTs without weakly-coupled UV descriptions, similar to the F-theory
example discussed above. It would no doubt be interesting to calculate $\Delta
S$ and $\Delta T$ in such situations.

\section*{Acknowledgements}

We thank T. Dumitrescu, M. Freytsis and A. Katz for helpful discussions, and
especially K. Intriligator and P. Langacker for helpful discussions and comments.
The work of JJH is supported by NSF grant PHY-1067976. The work of PK is
supported by DOE\ grant DE-FG02-92ER40704. The
work of BW is supported by the STFC\ Standard\ Grant ST/J000469/1
\textquotedblleft String Theory, Gauge Theory and Duality\textquotedblright.

%%%%%%%%%%%%%%%%%%%%%%%%%%%%%%%%%%%%%%%%%%%%%%%%%%%%%%%%%%%%%%%%%%%%%%%%%%%%%%%%%%%%%%%%%%%%%%%%%%%%%%%%%%%

\bibliographystyle{utphys}
\bibliography{PrecisionHiggs}

\providecommand{\href}[2]{#2}\begingroup\raggedright\begin{thebibliography}{10}

\bibitem{Aad:2012gk}
{\bfseries ATLAS Collaboration}, G.~Aad {\em et al.},
  ``{Observation of a New Particle in the Search for the Standard Model Higgs
  Boson with the ATLAS Detector at the LHC},''
  \href{http://dx.doi.org/10.1016/j.physletb.2012.08.020}{{\em Phys. Lett.}
  {\bfseries B716} (2012) 1--29},
\href{http://arxiv.org/abs/1207.7214}{{\ttfamily arXiv:1207.7214 [hep-ex]}}.
%%CITATION = ARXIV:1207.7214;%%.

\bibitem{Chatrchyan:2012gu}
{\bfseries CMS Collaboration}, S.~Chatrchyan {\em et al.},
  ``{Observation of a new boson at a mass of 125 GeV with the CMS experiment at
  the LHC},'' \href{http://dx.doi.org/10.1016/j.physletb.2012.08.021}{{\em
  Phys. Lett.} {\bfseries B716} (2012) 30--61},
\href{http://arxiv.org/abs/1207.7235}{{\ttfamily arXiv:1207.7235 [hep-ex]}}.
%%CITATION = ARXIV:1207.7235;%%.

\bibitem{Shifman:1979eb}
M.~A. Shifman, A.~I. Vainshtein, M.~B. Voloshin, and V.~I. Zakharov,
  ``{Low-Energy Theorems for Higgs Boson Couplings to Photons},''
{\em Sov. J. Nucl. Phys.} {\bfseries 30} (1979) 711--716.
%%CITATION = SJNCA,30,711;%%.

\bibitem{Peskin:1990zt}
M.~E. Peskin and T.~Takeuchi, ``{A New constraint on a strongly interacting
  Higgs sector},''
\href{http://dx.doi.org/10.1103/PhysRevLett.65.964}{{\em Phys. Rev. Lett.}
  {\bfseries 65} (1990) 964--967}.
%%CITATION = PRLTA,65,964;%%.

\bibitem{Peskin:1991sw}
M.~E. Peskin and T.~Takeuchi, ``{Estimation of oblique electroweak
  corrections},''
\href{http://dx.doi.org/10.1103/PhysRevD.46.381}{{\em Phys. Rev.} {\bfseries
  D46} (1992) 381--409}.
%%CITATION = PHRVA,D46,381;%%.

\bibitem{ErlerLangackerPDG}
J.~Erler and P.~Langacker, ``{Electroweak Model and Constraints on New
  Physics},''
\href{http://dx.doi.org/10.1103/PhysRevD.86.010001}{{\em in J. Beringer et al.
  (Particle Data Group), Phys. Rev.} {\bfseries D86} (2012) 010001}.
%%CITATION = PHRVA,D86,010001;%%.

\bibitem{ALEPH:2005ab}
{\bfseries ALEPH Collaboration, DELPHI Collaboration, L3 Collaboration, OPAL
  Collaboration, SLD Collaboration, LEP Electroweak Working Group, SLD
  Electroweak Group, SLD Heavy Flavour Group}, S.~Schael {\em et
  al.}, ``{Precision electroweak measurements on the $Z$ resonance},''
  \href{http://dx.doi.org/10.1016/j.physrep.2005.12.006}{{\em Phys. Rept.}
  {\bfseries 427} (2006) 257--454},
\href{http://arxiv.org/abs/hep-ex/0509008}{{\ttfamily arXiv:hep-ex/0509008}}.
%%CITATION = HEP-EX/0509008;%%.

\bibitem{Novikov:1983uc}
V.~A. Novikov, M.~A. Shifman, A.~I. Vainshtein, and V.~I. Zakharov, ``{Exact
  Gell-Mann-Low Function of Supersymmetric Yang-Mills Theories from Instanton
  Calculus},''
\href{http://dx.doi.org/10.1016/0550-3213(83)90338-3}{{\em Nucl. Phys.}
  {\bfseries B229} (1983) 381}.
%%CITATION = NUPHA,B229,381;%%.

\bibitem{Stancato:2008mp}
D.~Stancato and J.~Terning, ``{The Unhiggs},''
  \href{http://dx.doi.org/10.1088/1126-6708/2009/11/101}{{\em JHEP} {\bfseries
  0911} (2009) 101}, \href{http://arxiv.org/abs/0807.3961}{{\ttfamily
  arXiv:0807.3961 [hep-ph]}}.

\bibitem{Komargodski:2008ax}
Z.~Komargodski and N.~Seiberg, ``{$\mu$ and General Gauge Mediation},''
  \href{http://dx.doi.org/10.1088/1126-6708/2009/03/072}{{\em JHEP} {\bfseries
  0903} (2009) 072},
\href{http://arxiv.org/abs/0812.3900}{{\ttfamily arXiv:0812.3900 [hep-ph]}}.
%%CITATION = ARXIV:0812.3900;%%.

\bibitem{Azatov:2011ht}
A.~Azatov, J.~Galloway, and M.~A. Luty, ``{Superconformal Technicolor},''
  \href{http://dx.doi.org/10.1103/PhysRevLett.108.041802}{{\em Phys. Rev.
  Lett.} {\bfseries 108} (2012) 041802},
\href{http://arxiv.org/abs/1106.3346}{{\ttfamily arXiv:1106.3346 [hep-ph]}}.
%%CITATION = ARXIV:1106.3346;%%.

\bibitem{Azatov:2011ps}
A.~Azatov, J.~Galloway, and M.~A. Luty, ``{Superconformal Technicolor: Models
  and Phenomenology},''
  \href{http://dx.doi.org/10.1103/PhysRevD.85.015018}{{\em Phys. Rev.}
  {\bfseries D85} (2012) 015018},
\href{http://arxiv.org/abs/1106.4815}{{\ttfamily arXiv:1106.4815 [hep-ph]}}.
%%CITATION = ARXIV:1106.4815;%%.

\bibitem{Gherghetta:2011na}
T.~Gherghetta and A.~Pomarol, ``{A Distorted MSSM Higgs Sector from Low-Scale
  Strong Dynamics},'' \href{http://dx.doi.org/10.1007/JHEP12(2011)069}{{\em
  JHEP} {\bfseries 1112} (2011) 069},
\href{http://arxiv.org/abs/1107.4697}{{\ttfamily arXiv:1107.4697 [hep-ph]}}.
%%CITATION = ARXIV:1107.4697;%%.

\bibitem{DSSM}
J.~J. Heckman, P.~Kumar, C.~Vafa, and B.~Wecht, ``{Electroweak Symmetry
  Breaking in the DSSM},''
  \href{http://dx.doi.org/10.1007/JHEP01(2012)156}{{\em JHEP} {\bfseries 1201}
  (2012) 156},
\href{http://arxiv.org/abs/1108.3849}{{\ttfamily arXiv:1108.3849 [hep-ph]}}.
%%CITATION = ARXIV:1108.3849;%%.

\bibitem{HoloHiggs}
J.~J. Heckman, P.~Kumar, and B.~Wecht, ``{The Higgs as a Probe of
  Supersymmetric Extra Sectors},''
  \href{http://dx.doi.org/10.1007/JHEP07(2012)118}{{\em JHEP} {\bfseries 1207}
  (2012) 118},
\href{http://arxiv.org/abs/1204.3640}{{\ttfamily arXiv:1204.3640 [hep-ph]}}.
%%CITATION = ARXIV:1204.3640;%%.

\bibitem{Kitano:2012wv}
R.~Kitano, M.~A. Luty, and Y.~Nakai, ``{Partially Composite Higgs in
  Supersymmetry},'' \href{http://dx.doi.org/10.1007/JHEP08(2012)111}{{\em JHEP}
  {\bfseries 1208} (2012) 111},
\href{http://arxiv.org/abs/1206.4053}{{\ttfamily arXiv:1206.4053 [hep-ph]}}.
%%CITATION = ARXIV:1206.4053;%%.

\bibitem{Funparticles}
J.~J. Heckman and C.~Vafa, ``{An Exceptional Sector for F-theory GUTs},''
  \href{http://dx.doi.org/10.1103/PhysRevD.83.026006}{{\em Phys. Rev.}
  {\bfseries D83} (2011) 026006},
\href{http://arxiv.org/abs/1006.5459}{{\ttfamily arXiv:1006.5459 [hep-th]}}.
%%CITATION = 1006.5459;%%.

\bibitem{HVW}
J.~J. Heckman, C.~Vafa, and B.~Wecht, ``{The Conformal Sector of F-theory
  GUTs},'' \href{http://dx.doi.org/10.1007/JHEP07(2011)075}{{\em JHEP}
  {\bfseries 1107} (2011) 075},
  \href{http://arxiv.org/abs/1103.3287}{{\ttfamily arXiv:1103.3287 [hep-th]}}.

\bibitem{Intriligator:2003jj}
K.~A. Intriligator and B.~Wecht, ``{The exact superconformal R-symmetry
  maximizes $a$},'' \href{http://dx.doi.org/10.1016/S0550-3213(03)00459-0}{{\em
  Nucl. Phys.} {\bfseries B667} (2003) 183--200},
\href{http://arxiv.org/abs/hep-th/0304128}{{\ttfamily arXiv:hep-th/0304128}}.
%%CITATION = HEP-TH/0304128;%%.

\bibitem{Dugan:1991ck}
M.~J. Dugan and L.~Randall, ``{The Sign of $S$ from electroweak radiative
  corrections},''
\href{http://dx.doi.org/10.1016/0370-2693(91)90720-B}{{\em Phys. Lett.}
  {\bfseries B264} (1991) 154--160}.
%%CITATION = PHLTA,B264,154;%%.

\bibitem{Gates:1991uu}
E.~Gates and J.~Terning, ``{Negative contributions to S from Majorana
  particles},'' \href{http://dx.doi.org/10.1103/PhysRevLett.67.1840}{{\em Phys.
  Rev. Lett.} {\bfseries 67} (1991) 1840--1843}.

\bibitem{He:2001tp}
H.-J. He, N.~Polonsky, and S.~Su, ``{Extra Families, Higgs Spectrum and Oblique
  Corrections},'' \href{http://dx.doi.org/10.1103/PhysRevD.64.053004}{{\em
  Phys. Rev.} {\bfseries D64} (2001) 053004},
\href{http://arxiv.org/abs/hep-ph/0102144}{{\ttfamily arXiv:hep-ph/0102144}}.
%%CITATION = HEP-PH/0102144;%%.

\bibitem{Skiba:2010xn}
W.~Skiba, ``{TASI Lectures on Effective Field Theory and Precision Electroweak
  Measurements},''
\href{http://arxiv.org/abs/1006.2142}{{\ttfamily arXiv:1006.2142 [hep-ph]}}.
%%CITATION = ARXIV:1006.2142;%%.

\bibitem{GGMI}
P.~Meade, N.~Seiberg, and D.~Shih, ``{General Gauge Mediation},''
  \href{http://dx.doi.org/10.1143/PTPS.177.143}{{\em Prog. Theor. Phys. Suppl.}
  {\bfseries 177} (2009) 143--158},
\href{http://arxiv.org/abs/0801.3278}{{\ttfamily arXiv:0801.3278 [hep-ph]}}.
%%CITATION = 0801.3278;%%.

\bibitem{GGMII}
M.~Buican, P.~Meade, N.~Seiberg, and D.~Shih, ``{Exploring General Gauge
  Mediation},'' \href{http://dx.doi.org/10.1088/1126-6708/2009/03/016}{{\em
  JHEP} {\bfseries 03} (2009) 016},
\href{http://arxiv.org/abs/0812.3668}{{\ttfamily arXiv:0812.3668 [hep-ph]}}.
%%CITATION = 0812.3668;%%.

\bibitem{Dumitrescu:2011iu}
T.~T. Dumitrescu and N.~Seiberg, ``{Supercurrents and Brane Currents in Diverse
  Dimensions},'' \href{http://dx.doi.org/10.1007/JHEP07(2011)095}{{\em JHEP}
  {\bfseries 1107} (2011) 095},
\href{http://arxiv.org/abs/1106.0031}{{\ttfamily arXiv:1106.0031 [hep-th]}}.
%%CITATION = ARXIV:1106.0031;%%.

\bibitem{Fortin:2011nq}
J.-F. Fortin, K.~Intriligator, and A.~Stergiou, ``{Current OPEs in
  Superconformal Theories},''
  \href{http://dx.doi.org/10.1007/JHEP09(2011)071}{{\em JHEP} {\bfseries 09}
  (2011) 071},
\href{http://arxiv.org/abs/1107.1721}{{\ttfamily arXiv:1107.1721 [hep-th]}}.
%%CITATION = 1107.1721;%%.

\bibitem{Fortin:2011ad}
J.-F. Fortin, K.~Intriligator, and A.~Stergiou, ``{Superconformally Covariant
  OPE and General Gauge Mediation},''
  \href{http://dx.doi.org/10.1007/JHEP12(2011)064}{{\em JHEP} {\bfseries 12}
  (2011) 064},
\href{http://arxiv.org/abs/1109.4940}{{\ttfamily arXiv:1109.4940 [hep-th]}}.
%%CITATION = 1109.4940;%%.

\bibitem{Coleman:1973jx}
S.~R. Coleman and E.~J. Weinberg, ``{Radiative Corrections as the Origin of
  Spontaneous Symmetry Breaking},''
\href{http://dx.doi.org/10.1103/PhysRevD.7.1888}{{\em Phys. Rev.} {\bfseries
  D7} (1973) 1888--1910}.
%%CITATION = PHRVA,D7,1888;%%.

\bibitem{Mack:1975je}
G.~Mack, ``{All Unitary Ray Representations of the Conformal Group $SU(2,2)$
  with Positive Energy},''
\href{http://dx.doi.org/10.1007/BF01613145}{{\em Commun. Math. Phys.}
  {\bfseries 55} (1977) 1}.
%%CITATION = CMPHA,55,1;%%.

\bibitem{Grinstein:2008qk}
B.~Grinstein, K.~A. Intriligator, and I.~Z. Rothstein, ``{Comments on
  Unparticles},'' \href{http://dx.doi.org/10.1016/j.physletb.2008.03.020}{{\em
  Phys. Lett.} {\bfseries B662} (2008) 367--374},
\href{http://arxiv.org/abs/0801.1140}{{\ttfamily arXiv:0801.1140 [hep-ph]}}.
%%CITATION = ARXIV:0801.1140;%%.

\bibitem{Anselmi:1997am}
D.~Anselmi, D.~Z. Freedman, M.~T. Grisaru, and A.~A. Johansen,
  ``{Non-perturbative formulas for central functions of supersymmetric gauge
  theories},'' \href{http://dx.doi.org/10.1016/S0550-3213(98)00278-8}{{\em
  Nucl. Phys.} {\bfseries B526} (1998) 543--571},
\href{http://arxiv.org/abs/hep-th/9708042}{{\ttfamily arXiv:hep-th/9708042}}.
%%CITATION = HEP-TH/9708042;%%.

\bibitem{Anselmi:1997ys}
D.~Anselmi, J.~Erlich, D.~Freedman, and A.~Johansen, ``{Positivity constraints
  on anomalies in supersymmetric gauge theories},''
  \href{http://dx.doi.org/10.1103/PhysRevD.57.7570}{{\em Phys. Rev.} {\bfseries
  D57} (1998) 7570--7588},
\href{http://arxiv.org/abs/hep-th/9711035}{{\ttfamily arXiv:hep-th/9711035}}.
%%CITATION = HEP-TH/9711035;%%.

\bibitem{Bellantoni:2012ag}
L.~Bellantoni, J.~Erler, J.~J. Heckman, and E.~Ramirez-Homs, ``{Masses of a
  Fourth Generation with Two Higgs Doublets},''
  \href{http://dx.doi.org/10.1103/PhysRevD.86.034022}{{\em Phys. Rev.}
  {\bfseries D86} (2012) 034022},
\href{http://arxiv.org/abs/1205.5580}{{\ttfamily arXiv:1205.5580 [hep-ph]}}.
%%CITATION = ARXIV:1205.5580;%%.

\bibitem{Hook:2012fp}
A.~Hook, ``{A test for emergent dynamics},''
  \href{http://dx.doi.org/10.1007/JHEP07(2012)040}{{\em JHEP} {\bfseries 1207}
  (2012) 040},
\href{http://arxiv.org/abs/1204.4466}{{\ttfamily arXiv:1204.4466 [hep-th]}}.
%%CITATION = ARXIV:1204.4466;%%.

\bibitem{Amariti:2012wc}
A.~Amariti and K.~Intriligator, ``{$\Delta a$ curiosities in some 4d susy RG
  flows},'' \href{http://dx.doi.org/10.1007/JHEP11(2012)108}{{\em JHEP}
  {\bfseries 1211} (2012) 108},
\href{http://arxiv.org/abs/1209.4311}{{\ttfamily arXiv:1209.4311 [hep-th]}}.
%%CITATION = ARXIV:1209.4311;%%.

\bibitem{HVLHC}
J.~J. Heckman and C.~Vafa, ``{From F-theory GUTs to the LHC},''
\href{http://arxiv.org/abs/0809.3452}{{\ttfamily arXiv:0809.3452 [hep-ph]}}.
%%CITATION = 0809.3452;%%.

\bibitem{Heckman:2010bq}
J.~J. Heckman, ``{Particle Physics Implications of F-theory},'' {\em Ann. Rev.
  Nuc. Part. Sci.} {\bfseries 60} (2010) 237,
\href{http://arxiv.org/abs/1001.0577}{{\ttfamily arXiv:1001.0577 [hep-th]}}.
%%CITATION = 1001.0577;%%.

\bibitem{Weigand:2010wm}
T.~Weigand, ``{Lectures on F-theory compactifications and model building},''
  {\em Class. Quant. Grav.} {\bfseries 27} (2010) 214004,
\href{http://arxiv.org/abs/arXiv:1009.3497 [hep-th]}{{\ttfamily arXiv:1009.3497
  [hep-th]}}.
%%CITATION = 1009.3497;%%.

\bibitem{Maharana:2012tu}
A.~Maharana and E.~Palti, ``{Models of Particle Physics from Type IIB String
  Theory and F-theory: A Review},''
\href{http://arxiv.org/abs/1212.0555}{{\ttfamily arXiv:1212.0555 [hep-th]}}.
%%CITATION = ARXIV:1212.0555;%%.

\bibitem{MNI}
J.~A. Minahan and D.~Nemeschansky, ``{An $\mathcal{N} = 2$ superconformal fixed
  point with $E_6$ global symmetry},''
  \href{http://dx.doi.org/10.1016/S0550-3213(96)00552-4}{{\em Nucl. Phys.}
  {\bfseries B482} (1996) 142--152},
\href{http://arxiv.org/abs/hep-th/9608047}{{\ttfamily arXiv:hep-th/9608047}}.
%%CITATION = HEP-TH/9608047;%%.

\bibitem{MNII}
J.~A. Minahan and D.~Nemeschansky, ``{Superconformal fixed points with $E_n$
  global symmetry},''
  \href{http://dx.doi.org/10.1016/S0550-3213(97)00039-4}{{\em Nucl. Phys.}
  {\bfseries B489} (1997) 24--46},
\href{http://arxiv.org/abs/hep-th/9610076}{{\ttfamily arXiv:hep-th/9610076}}.
%%CITATION = HEP-TH/9610076;%%.

\bibitem{FCFT}
J.~J. Heckman, Y.~Tachikawa, C.~Vafa, and B.~Wecht, ``{$\mathcal{N} = 1$ SCFTs
  from Brane Monodromy},''
  \href{http://dx.doi.org/10.1007/JHEP11(2010)132}{{\em JHEP} {\bfseries 11}
  (2010) 132},
\href{http://arxiv.org/abs/1009.0017}{{\ttfamily arXiv:1009.0017 [hep-th]}}.
%%CITATION = 1009.0017;%%.

\bibitem{D3gen}
J.~J. Heckman and S.-J. Rey, ``{Baryon and Dark Matter Genesis from Strongly
  Coupled Strings},'' \href{http://dx.doi.org/10.1007/JHEP06(2011)120}{{\em
  JHEP} {\bfseries 06} (2011) 120},
\href{http://arxiv.org/abs/1102.5346}{{\ttfamily arXiv:1102.5346 [hep-th]}}.
%%CITATION = 1102.5346;%%.

\end{thebibliography}\endgroup

\end{document}